\documentclass[epj]{svjour}        
\usepackage{amsmath,bm,graphicx,psfrag,layout,float}
\usepackage{amsbsy,amssymb}       
\begin{document}                   
\title{The Study of Goldstone Modes in $\nu$=2 Bilayer Quantum Hall Systems }  
\author{Y. Hama\inst{1,2}, Y. Hidaka\inst{2}, 
G. Tsitsishvili\inst{3}, 
\and Z. F. Ezawa\inst{4}}                   

\institute{Department of Physics, The University of Tokyo, 113-0033, Tokyo, Japan, \and
Theoretical Research Division, Nishina Center,     
RIKEN, Wako 351-0198, Japan, \and
Department of Physics, Tbilisi State University, Tbilisi 0128, Georgia \and
Advanced Meson Science Laborary, Nishina Center, RIKEN, Wako 351-0198, Japan}  
\date{Received: date / Revised version: date}
\abstract{
At the filling factor $\nu$=2, the bilayer quantum Hall system has three phases, 
the spin-ferromagnet phase, the spin singlet phase and the canted antiferromagnet (CAF) phase, 
depending on the relative strength between the Zeeman energy and  interlayer tunneling energy.
We present a systematic method to derive the effective Hamiltonian for the Goldstone modes in these three phases.
We then investigate the dispersion relations and the coherence lengths of the Goldstone modes.
To explore a possible emergence of the interlayer phase coherence,
we analyze the dispersion relations in the zero tunneling energy  limit.
We find one gapless mode  with the linear dispersion relation in the CAF phase. 
}
%
\authorrunning{ Y. Hama {\it{et al}}. } 
\titlerunning{The Study of Goldstone Modes in $\nu$=2 Bilayer Quantum Hall System}
\maketitle
\section{Introduction} 
\label{intro}

In the bilayer quantum Hall (QH) system, a rich physics emerges
by the interplay between the spin and the layer (pseudospin) degrees of freedom%
\cite{Ezawa:2008ae,S. Das Sarma1}.    
For instance,  at the filling factor $\nu=1$, there arises uniquely
the spin-ferro
magnet 
and pseudospin-ferromagnet phase, 
showing various intralayer and interlayer coherent phenomena. 
On the other hand, the phases arising at $\nu=2$ are quite nontrivial.  
According to the one-body picture 
we expect to have two phases depending on the relative strength between the Zeeman gap $\Delta_{\text{Z}}$ and 
the tunneling gap $\Delta_{\text{SAS}}$. 
One is the spin-ferromagnet and pseudospin-singlet phase (abridged as the spin phase) for $\Delta_{\text{Z}}>\Delta_{\text{SAS}}$;
 the other is the spin-singlet and pseudospin ferromagnet phase (abridged as the pseudospin phase) for $\Delta_{\text{SAS}}>\Delta_{\text{Z}}$. 
Instead, an intermediate phase, a canted antiferromagnetic phase  (abridged as the CAF phase) emerges.
This is a novel phase where the spin  direction is canted and makes antiferromagnetic correlations
between  the two layers\cite{S. Das Sarma2,S. Das Sarma3}.  
Das Sarma {\it{et al}}. obtained the phase diagram in the $\Delta_{\text{SAS}}-d$ plane based on time-dependent Hartree-Fock  analysis, where $d$ is the layer separation\cite{S. Das Sarma2,S. Das Sarma3}.
Later on, an effective spin theory, a  Hartree-Fock-Bogoliubov approximation and an exact diagonalization  
study were employed to improve the phase diagram%
\cite{S. Das Sarma4,Yang1,Shimoda1,MacDonald1}. 
Effects of the density imbalance on the CAF were also discussed\cite{MacDonald2,S. Das Sarma5}. 
    
The first experimental indication  of the CAF phase was given by  
inelastic light scattering spectroscopy\cite{Pellegrini1}. 
They also have observed softening signals indicating second-order phase transitions\cite{Pellegrini2}. 
Subsequently, an unambiguous evidence of the CAF phase was obtained through capacitance spectroscopy 
as well as 
magnetotransport measurements 
\cite{khrapai1,S. J. Geer1,Sawada1,Sawada2,Sawada3,Sawada4}. 

The ground state structure of the $\nu =2$ bilayer QH system
has been investigated based on the SU(4) formalism
\cite{Ezawa:2005xi,Ezawa:2001dn,hasebe1,Ezawa:2003sr,Tsitsishvili:2005gv,Ezawa:2005cj}. 
The expectation values of the SU(4) isospin operators are the order parameters,
in terms of which an anisotropic SU(4) nonlinear sigma model has been derived 
to describe low-energy coherent phenomena\cite{Ezawa:2005xi}.  
However, the effective Hamiltonian for the Goldstone modes
has not been derived. Though there are some results with the use of 
Grassmannian fields in the spin and pseudospin phases, no attempts have been made
in the CAF phase. On the other hand, experimentally, a role of a Goldstone
mode has been suggested by nuclear magnetic resonance\cite{Kumada1}
in the CAF phase. 
 
In this paper we develop a generic formalism to determine the symmetry
breaking pattern and to derive the effective Hamiltonian for the Goldstone
modes in the three phases of the $\nu =2$ bilayer QH system. The symmetry
breaking pattern reads
\begin{equation}
\text{SU}(4)\rightarrow \text{U}(1)\otimes \text{SU}(2)\otimes \text{SU}(2),
\end{equation}
and there appear eight Goldstone modes in each phase. The corresponding
Goldstone modes in the two phases match smoothly at the phase boundary. All
the modes are actually gapped except along the phase boundaries due to explicit
symmetry breaking terms. It is important if gapless modes emerge in the limit 
$\Delta _{\text{Z}}\rightarrow 0$ or $\Delta _{\text{SAS}}\rightarrow 0$,
where the spin coherence or the interlayer coherence is enhanced. Gapless
modes are genuine Goldstone modes associated with spontaneous symmetry
breaking. Naturally we have gapless modes in the spin phase as 
$\Delta _{\text{Z}}\rightarrow 0$ and in the pseudospin phase 
as $\Delta _{\text{SAS}}\rightarrow 0$. 
It is intriguing that we find one gapless mode with the linear dispersion relation
in the CAF phase as $\Delta _{\text{SAS}}\rightarrow 0$.

This paper is organized as follows. In Sec. \ref{sec:II},  
we review  the Coulomb interaction of the bilayer QH system projected to the lowest Landau level (LLL) and  the SU(4) effective Hamiltonian after making the derivative expansion.
We also review the ground state structure in the three phases.
In Sec. \ref{sec:III}, which is the main part of this paper, we develop a unified formalism to derive
the  effective Hamiltonian for the Goldstone modes.
Then we discuss the SU(4) symmetry breaking pattern and the Goldstone mode spectrum, 
such as the dispersion relations and the coherence length in each phase. 
In particular, for the investigation of the CAF phase, we find it useful to introduce two convenient coordinates of SU(4) group space, 
the s-coordinate and the p-coordinate.
We study the dispersions and the coherence length in the limit $\Delta_{\text{SAS}}\rightarrow0$,
to explore a possible emergence of the interlayer phase coherence in the CAF phase.
Remarkably, we find one coherent mode whose coherence length diverges.  
Section \ref{sec:IV} is devoted to discussion.

\section{The SU(4) Effective Hamiltonian and the ground state structure}

\label{sec:II}

Electrons in a plane perform cyclotron motion under perpendicular magnetic
field $B_{\perp }$ and create Landau levels. The number of flux quanta
passing through the system is $N_{\Phi }\equiv B_{\perp }S/\Phi _{\text{D}}$, 
where $S$ is the area of the system and $\Phi _{\text{D}}=2\pi \hbar /e$
is the flux quantum. There are $N_{\Phi }$ Landau sites per one Landau
level, each of which is associated with one flux quantum and occupies an
area $S/N_{\Phi }=2\pi \ell _{B}^{2}$, with the magnetic length 
$\ell _{B}=\sqrt{\hbar /eB_{\perp }}$.

In the bilayer system an electron has two types of indices, the spin index 
$(\uparrow ,\downarrow )$ and the layer index $(\text{f},\text{b})$. They can
be incorporated in 4 types of isospin index 
$\alpha =$ f$\uparrow $,f$\downarrow $,b$\uparrow $,b$\downarrow $. 
One Landau site may contain four
electrons. The filling factor is $\nu =N/N_{\Phi }$ with $N$ the total
number of electrons.

We explore the physics of electrons confined to the LLL,
where the electron position is specified solely by the guiding center 
$\boldsymbol{X}=(X,Y)$, whose $X$ and $Y$ components are noncommutative, 
\begin{equation}
\lbrack X,Y]=-i\ell _{B}^{2}.  \label{AlgebGC}
\end{equation}
The equations of motion follow from this noncommutative relation rather than the kinetic term for electrons confined within the LLL.
In order to derive the effective Hamiltonian, it is convenient to represent the noncommutative relation with the use of the Fock states,
\begin{equation}
|n\rangle =\frac{1}{\sqrt{n!}}(b^{\dag })^{n}|0\rangle ,\quad n=0,1,2,\cdots
,\quad b|0\rangle =0,  \label{LandaSite}
\end{equation}
where $b$ and $b^{\dag }$ are the ladder operators, 
\begin{equation}
b=\frac{1}{\sqrt{2}\ell _{B}}(X-iY),\qquad b^{\dag }=\frac{1}{\sqrt{2}\ell_{B}}(X+iY),  \label{WeylXb}
\end{equation}
obeying $[b,b^{\dag }]=1$.
Although the Fock states correspond to the Landau sites in the symmetric gauge,
the resulting effective Hamiltonian is independent of the representation
we have chosen.

We expand the electron field operator by a complete set of one-body wave
functions $\varphi _{n}(\boldsymbol{x})=\langle \boldsymbol{x}|n\rangle $ in the LLL, 
\begin{equation}
\psi _{\alpha }(\boldsymbol{x})\equiv \sum_{n=1}^{N_{\Phi }}c_{\alpha}(n)\varphi _{n}(\boldsymbol{x}),  \label{electronfield}
\end{equation}
where $c_{\alpha }(n)$ is the annihilation operator at the Landau site 
$|n\rangle $ with $\alpha =$ f$\uparrow $,f$\downarrow $,b$\uparrow $,b$\downarrow $. 
The operators $c_{\alpha }(m),c_{\beta }^{\dagger }(n)$
satisfy the standard anticommutation relations,
\begin{align}
\{c_{\alpha }(m),c_{\beta }^{\dagger }(n)\}& =\delta _{mn}\delta _{\alpha\beta },  \notag \\
\{c_{\alpha }(m),c_{\beta }(n)\}& =\{c_{\alpha }^{\dagger }(m),c_{\beta}^{\dagger }(n)\}=0.  
\end{align}
The electron field $\psi _{\alpha }(\boldsymbol{x})$ has four components, and
the bilayer system possesses the underlying algebra SU$(4)$ with having the
subalgebra $\text{SU}_{\text{spin}}(2)\times \text{SU}_{\text{ppin}}(2)$. 
We denote the three generators of the $\text{SU}_{\text{spin}}(2)$ by $\tau
_{a}^{\text{spin}}$, and those of $\text{SU}_{\text{ppin}}(2)$ by $\tau
_{a}^{\text{ppin}}$. There are remaining nine generators 
$\tau _{a}^{\text{spin}}\tau _{b}^{\text{ppin}}$. Their explicit form is given in Apendix A.

All the physical operators required for the description of the system are
constructed as the bilinear combinations of $\psi (\boldsymbol{x})$ and 
$\psi^{\dagger }(\boldsymbol{x})$. They are 16 density operators 
\begin{align}
\rho (\boldsymbol{x})& =\psi ^{\dagger }(\boldsymbol{x})\psi (\boldsymbol{x}),  \notag \\
S_{a}(\boldsymbol{x})& =\frac{1}{2}\psi ^{\dagger }(\boldsymbol{x})\tau _{a}^{\text{spin}}
\psi (\boldsymbol{x}),  \notag \\
P_{a}(\boldsymbol{x})& =\frac{1}{2}\psi ^{\dagger }(\boldsymbol{x})\tau _{a}^{\text{ppin}}
\psi (\boldsymbol{x}),  \notag \\
R_{ab}(\boldsymbol{x})& =\frac{1}{2}\psi ^{\dagger }(\boldsymbol{x})\tau _{a}^{\text{spin}}
\tau _{b}^{\text{ppin}}\psi (\boldsymbol{x}),  \label{su4isospin1}
\end{align}
where $S_{a}$ describes the total spin, $2P_{z}$ measures the
electron-density difference between the two layers. The operator $R_{ab}$
transforms as a spin under $\text{SU}_{\text{spin}}(2)$ and as a pseudospin
under $\text{SU}_{\text{ppin}}(2)$. 

The kinetic Hamiltonian is quenched, since the kinetic energy is common to
all states in the LLL. The Coulomb interaction is decomposed into the
SU(4)-invariant and SU(4)-noninvariant terms 
\begin{align}
H_{\text{C}}^{+}& =\frac{1}{2}\int d^{2}xd^{2}yV^{+}(\boldsymbol{x}-\boldsymbol{y})
\rho (\boldsymbol{x})\rho (\boldsymbol{y}),  \label{coulomb1} \\
H_{\text{C}}^{-}& =2\int d^{2}xd^{2}yV^{-}(\boldsymbol{x}-\boldsymbol{y})P_{z}(\boldsymbol{x})
P_{z}(\boldsymbol{y}),  \label{coulomb2}
\end{align}
where 
\begin{equation}
V^{\pm }(\boldsymbol{x})=\frac{e^{2}}{8\pi \epsilon }\left( \frac{1}{|\boldsymbol{x}|}
\pm \frac{1}{\sqrt{|\boldsymbol{x}|^{2}+d^{2}}}\right),
\end{equation}
with the layer separation $d$. The tunneling and bias terms are summarized
into the pseudo-Zeeman term. Combining the Zeeman and pseudo-Zeeman terms 
we have 
\begin{equation}
H_{\text{ZpZ}}=-\int d^{2}x(\Delta _{\text{Z}}S_{z}+\Delta _{\text{SAS}}P_{x}+\Delta _{\text{bias}}P_{z}),
\end{equation}
with the Zeeman gap $\Delta _{\text{Z}}$, the tunneling gap $\Delta _{\text{SAS}}$, 
and the bias voltage $\Delta _{\text{bias}}=eV_{\text{bias}}$.

The total Hamiltonian is 
\begin{equation}
H=H_{\text{C}}^{+}+H_{\text{C}}^{-}+H_{\text{ZpZ}}.
\end{equation}
We investigate the regime where the SU(4) invariant \\
Coulomb term $H_{C}^{+}$
dominates all other interactions. Note that the SU(4)-noninvariant terms
vanish in the limit $d$, $\Delta _{\text{Z}}$, $\Delta _{\text{SAS}}$, $\Delta _{\text{bias}}\rightarrow 0$.

We project the density operators (\ref{su4isospin1}) to the LLL by
substituting the field operator (\ref{electronfield}) into them. A typical
density operator reads
\begin{equation}
R_{ab}(\boldsymbol{p})=e^{-\ell _{B}^{2}\boldsymbol{p}^{2}/4}\hat{R}_{ab}(\boldsymbol{p}),
\end{equation}
in the momentum space, with
\begin{equation}
\hat{R}_{ab}(\boldsymbol{p})=\frac{1}{4\pi }\sum_{mn}
\langle n|e^{-i\boldsymbol{pX}}|m\rangle c^{\dag }(n)\tau _{a}^{\text{spin}}
\tau _{b}^{\text{ppin}}c(m),  \label{BareDensiR}
\end{equation}
where $c(m)$ is the $4$-component vector made of the operators $c_{\alpha}(m)$.

What are observed experimentally are the classical densities, which are
expectation values such as $\hat{\rho}^{\text{cl}}(\boldsymbol{p})=\langle 
\mathfrak{S}|\hat{\rho}(\boldsymbol{p})|\mathfrak{S}\rangle $, 
where $|\mathfrak{S}\rangle $ represents a generic state in the LLL.   
The Coulomb Hamiltonian governing the classical densities are given by\cite{Ezawa:2003sr} 
\begin{align}
H^{\text{eff}}& =\pi \int d^{2}pV_{D}^{+}(\boldsymbol{p})\hat{\rho}^{\text{cl}}
(-\boldsymbol{p})\hat{\rho}^{\text{cl}}(\boldsymbol{p})  \notag \\
& +4\pi \int d^{2}pV_{D}^{-}(\boldsymbol{p})\hat{P}_{z}^{\text{cl}}(-\boldsymbol{p})
\hat{P}_{z}^{\text{cl}}(\boldsymbol{p})  \notag \\
& -\frac{\pi }{2}\int d^{2}pV_{X}^{d}(\boldsymbol{p})[\hat{S}_{a}^{\text{cl}}
(-\boldsymbol{p})\hat{S}_{a}^{\text{cl}}(\boldsymbol{p})+\hat{P}_{a}^{\text{cl}}
(-\boldsymbol{p})\hat{P}_{a}^{\text{cl}}(\boldsymbol{p})  \notag \\
& +\hat{R}_{ab}^{\text{cl}}(-\boldsymbol{p})\hat{R}_{ab}^{\text{cl}}(\boldsymbol{p})]-\pi 
\int d^{2}pV_{X}^{-}(\boldsymbol{p})[\hat{S}_{a}^{\text{cl}}(-\boldsymbol{p})
\hat{S}_{a}^{\text{cl}}(\boldsymbol{p})  \notag \\
& +\hat{P}_{z}^{\text{cl}}(-\boldsymbol{p})\hat{P}_{z}^{\text{cl}}(\boldsymbol{p})+
\hat{R}_{az}^{\text{cl}}(-\boldsymbol{p})\hat{R}_{az}^{\text{cl}}(\boldsymbol{p})] 
\notag \\
& -\frac{\pi }{8}\int d^{2}pV_{X}(\boldsymbol{p})\hat{\rho}^{\text{cl}}(-\boldsymbol{p})
\hat{\rho}^{\text{cl}}(\boldsymbol{p}),
\end{align}
where $V_{D}$ and $V_{X}$ are the direct and exchange Coulomb potentials,
respectively, 
\begin{align}
V_{D}(\boldsymbol{p})& =\frac{e^{2}}{4\pi \epsilon |\boldsymbol{p}|}e^{-\ell _{B}^{2}
\boldsymbol{p}^{2}/2},  \notag \\
V_{X}(\boldsymbol{p})& =\frac{\sqrt{2\pi }e^{2}\ell_{B}}{4\pi \epsilon }I_{0}
(\ell_{B}^{2}\boldsymbol{p}^{2}/4)e^{-\ell _{B}^{2}\boldsymbol{p}^{2}/4},
\end{align}
with  $V_{X}=V_{X}^{+}+V_{X}^{-},\quad V_{X}^{d}=V_{X}^{+}-V_{X}^{-}$, and
\begin{align}
V_{D}^{\pm }(\boldsymbol{p})& =\frac{e^{2}}{8\pi \epsilon |\boldsymbol{p}|}
\left( 1\pm e^{-|\boldsymbol{p}|d}\right) e^{-\ell _{B}^{2}\boldsymbol{p}^{2}/2},  \notag \\
V_{X}^{\pm }(\boldsymbol{p})& =\frac{\sqrt{2\pi }e^{2}\ell _{B}}{8\pi\epsilon }
I_{0}(\ell _{B}^{2}\boldsymbol{p}^{2}/4)e^{-\ell _{B}^{2}
\boldsymbol{p}^{2}/4}  \notag \\
& \pm \frac{e^{2}\ell _{B}^{2}}{4\pi \epsilon }\int_{0}^{\infty }dke^{-\frac{1}{2}
\ell _{B}^{2}k^{2}-kd}J_{0}(\ell _{B}^{2}|\boldsymbol{p}|k).
\end{align}%
Here, $I_{0}(x)$ is the modified Bessel function, and $J_{0}(x)$ is the
Bessel function of the first kind. We comment that a similar Hamiltonian has
been derived based on the Schwinger boson mean-field theory\cite{MacDonald3}.

Since the exchange interaction $V^{\pm }(\boldsymbol{p})$ is short ranged, it is
a good approximation to make the derivative expansion, or equivalently, the
momentum expansion. 
We may set $\hat{\rho}^{\text{cl}}(\boldsymbol{p})=\rho _{0}$, 
$\hat{S}_{a}^{\text{cl}}(\boldsymbol{p})=\rho _{\Phi }\mathcal{S}_{a}(\boldsymbol{p})$,
$\hat{P}_{a}^{\text{cl}}(\boldsymbol{p})=\rho _{\Phi }\mathcal{P}_{a}(\boldsymbol{p})$, 
and $\hat{R}_{ab}^{\text{cl}}(\boldsymbol{p})=\rho _{\Phi }\mathcal{R}_{ab}(\boldsymbol{p})$ 
for the study of Goldstone modes. Taking the nontrivial lowest order
terms in the derivative expansion, we obtain the SU(4) effective Hamiltonian
density 
\begin{align} 
{\mathcal{H}}^{\text{eff}}& =J_{s}^{d}\left( \sum (\partial _{k}
\mathcal{S}_{a})^{2}+(\partial _{k}\mathcal{P}_{a})^{2}
+(\partial _{k}\mathcal{R}_{ab})^{2}\right)   \notag \\
& +2J_{s}^{-}\left( \sum (\partial _{k}\mathcal{S}_{a})^{2}+(\partial _{k}
\mathcal{P}_{z})^{2}+(\partial _{k}\mathcal{R}_{az})^{2}\right)   \notag \\
& +\rho _{\phi }[\epsilon _{\text{cap}}(\mathcal{P}_{z})^{2}-2\epsilon_{X}^{-}
\left( \sum (\mathcal{S}_{a})^{2}
+(\mathcal{R}_{az})^{2}\right)   \notag \\
& -(\epsilon _{X}^{+}-\epsilon _{X}^{-})(\sum \left( \mathcal{S}_{a})^{2}
+(\mathcal{P}_{a})^{2}+(\mathcal{R}_{ab})^{2}\right)   \notag \\
& -(\Delta _{\text{Z}}\mathcal{S}_{z}+\Delta _{\text{SAS}}\mathcal{P}_{x}
+\Delta _{\text{bias}}\mathcal{P}_{z})-(\epsilon _{X}^{+}+\epsilon _{X}^{-})],
\label{su4effectivehamiltonian1}
\end{align}
where $\rho_{\Phi }=\rho _{0}/\nu $ is the density of states, and
\begin{align}
J_s&=\frac{1}{16\sqrt{2\pi}}E^0_{\text{C}},\notag\\ 
J^d_s&=J_s \left[
-\sqrt{\frac{2}{\pi}}\frac{d}{\ell _{B}}+
\left(
1+\frac{d^2}{\ell _{B}^2}e^{d^2/2\ell _{B}^2}\text{erfc}\left(d/\sqrt{2}\ell _{B}  \right)
\right)
\right],\notag\\
J_{s}^{\pm }&=\frac{1}{2}(J_{s}\pm J_{s}^{d}),\notag\\
\epsilon_X&=\frac{1}{2}\sqrt{\frac{\pi}{2}}E^0_{\text{C}}, \quad
\epsilon_X^\pm=\frac{1}{2}\left[ 
1\pm  e^{d^2/2\ell _{B}^2}\text{erfc}\left(d/\sqrt{2}\ell _{B} \right)
\right]\epsilon_X,\notag\\ 
\epsilon _{D}^{-}&=\frac{d}{4\ell _{B}}E^0_\text{C}, \quad 
\epsilon _{\text{cap}}=4\epsilon _{D}^{-}-2\epsilon _{X}^{-}, 
\end{align}
with
\begin{equation}
E^0_{\text{C}}=\frac{e^2}{4\pi\epsilon \ell _{B}}.  
\end{equation}
This Hamiltonian is valid at $\nu =1, 2$ and $3$.

It is to be remarked that all potential terms vanish in the SU(4) invariant
limit, where perturbative excitations are gapless. They are the
Goldstone modes associated with spontaneous breaking of the SU(4)
symmetry. There are eight Goldstone modes, as we shall show in Section \ref{sec:III}. 
They get gapped in the actual system, since the SU(4) symmetry is
explicitly broken. Nevertheless we call them the Goldstone modes.

The ground state is obtained by minimizing the effective Hamiltonian 
(\ref{su4effectivehamiltonian1}) for homogeneous configurations of the classical
densities. The order parameters are the classical densities for the ground
state. It has been shown\cite{Ezawa:2005xi} at $\nu =2$ that they are given
in terms of two parameters $\alpha $ and $\beta $ as 
\begin{align} 
\mathcal{S}_{z}^{0}& =\frac{\Delta _{\text{Z}}}{\Delta _{0}}(1-\alpha ^{2})
\sqrt{1-\beta ^{2}},  \notag \\
\mathcal{P}_{x}^{0}& =\frac{\Delta _{\text{SAS}}}{\Delta _{0}}\alpha ^{2}
\sqrt{1-\beta ^{2}},\ \ \mathcal{P}_{z}^{0}=\frac{\Delta _{\text{SAS}}}
{\Delta _{0}}\alpha ^{2}\beta ,  \notag \\
\mathcal{R}_{xx}^{0}& =-\frac{\Delta _{\text{SAS}}}
{\Delta _{0}}\alpha \sqrt{1-\alpha ^{2}}\beta,  \notag \\
\mathcal{R}_{yy}^{0}& =-\frac{\Delta _{\text{Z}}}{\Delta _{0}}
\alpha \sqrt{1-\alpha ^{2}}\sqrt{1-\beta ^{2}},  \notag \\
\mathcal{R}_{xz}^{0}& =\frac{\Delta _{\text{SAS}}}
{\Delta _{0}}\alpha \sqrt{1-\alpha ^{2}}\sqrt{1-\beta ^{2}},
\label{orderparameter1}
\end{align}
with all others being zero.
The parameters $\alpha $ and $\beta $, satisfying $|\alpha |\leq 1$
and $|\beta |\leq 1$, are determined by the variational equations as 
\begin{align} 
\Delta_{\text{Z}}^{2} &=\frac{\Delta _{\text{SAS}}^{2}}{1-\beta ^{2}}
-\frac{4\epsilon _{X}^{-}\left( \Delta _{0}^{2}-\beta ^{2}
\Delta _{\text{SAS}}^{2}\right) }{\Delta _{0}\sqrt{1-\beta ^{2}}}, \label{vareq1}  \\ 
\frac{\Delta _{\text{bias}}}{\beta \Delta _{\text{SAS}}} &=
\frac{4\left(\epsilon _{X}^{-}+2\alpha ^{2}(\epsilon _{\text{D}}^{-}
-\epsilon_{X}^{-})\right) }{\Delta _{0}}+\frac{1}{\sqrt{1-\beta ^{2}}},
\label{vareq2}
\end{align} 
where 
\begin{equation}
\Delta _{0}=\sqrt{\Delta _{\text{SAS}}^{2}\alpha ^{2}
+\Delta_{\text{Z}}^{2}(1-\alpha ^{2})(1-\beta ^{2})}.  \label{EqD0}
\end{equation}
As a physical variable it is more convenient to use the imbalance parameter
defined by
\begin{equation}
\sigma _{0}\equiv \mathcal{P}_{z}^{0}=\frac{\Delta _{\text{SAS}}}
{\Delta _{0}}\alpha ^{2}\beta,   \label{ImbalParam}
\end{equation}
instead of the bias voltage $\Delta _{\text{bias}}$. 
This is possible in the pseudospin and CAF phases. 
The bilayer system is
balanced at $\sigma _{0}=0$, while all electrons are in the front layer at 
$\sigma _{0}=1$, and in the back layer at $\sigma _{0}=-1$. 

There are three phases in the bilayer QH system at $\nu =2$. We discuss them
in terms of $\alpha $ and $\beta $.

First, when $\alpha =0$, it follows that $\mathcal{S}_{z}^{0}=1$, 
$\mathcal{P}_{a}^{0}=\mathcal{R}_{ab}^{0}=0$, since $\Delta _{0}=\Delta_{\text{Z}}\sqrt{1-\beta ^{2}}$.
Note that $\beta $ disappears from all formulas in \eqref{orderparameter1}.
This is the spin phase, which is characterized by the fact that the isospin
is fully polarized into the spin direction with  
\begin{equation}
\mathcal{S}_{z}^{0}=1,
\end{equation}
and all others being zero. The spins in both layers point to the positive $z$
axis due to the Zeeman effect.

Second, when $\alpha =1$, it follows that $\mathcal{S}_{z}^{0}=0$ 
and $(\mathcal{P}_{x}^{0})^{2}+(\mathcal{P}_{z}^{0})^{2}=1$. This is the pseudospin
phase, which is characterized by the fact that the isospin is fully
polarized into the pseudospin direction with 
\begin{equation}
\mathcal{P}_{x}^{0}=\sqrt{1-\beta ^{2}},\qquad \mathcal{P}_{z}^{0}=\beta
=\sigma _{0},  \label{ppinorderparameters1}
\end{equation}
and all the others being zero. 

For intermediate values of $\alpha $ ($0<\alpha <1$), not only the spin and
pseudospin but also some components of the residual spin are nonvanishing,
where we may control the density imbalance by applying a bias voltage as in
the pseudospin phase. It follows from \eqref{orderparameter1} that, as the system
goes away from the spin phase $(\alpha =0)$, the spins begin to cant
coherently and make antiferromagnetic correlations between the two layers.
Hence it is called the canted antiferromagnetic phase.

The interlayer phase coherence is an intriguing phenomenon in the bilayer QH system\cite{Ezawa:2008ae}.
Since it is enhanced in the limit $\Delta _{\text{SAS}}\rightarrow 0$,
it is worthwhile to investigate the effective Hamiltonian in this limit. 
We need to know how the parameters 
$\alpha $ and $\beta $ are expressed in terms of the physical variables. Form 
\eqref{vareq1} it is trivial to see that $(1-\beta ^{2})/\Delta _{\text{SAS}}^{2}=O(1)$. 
Up to the order $O(\Delta _{\text{SAS}}^{2})$, \eqref{vareq2}
is reduced to 
\begin{equation}
\left( \Delta _{\text{Z}}^{2}-\frac{\Delta _{\text{SAS}}^{2}}{1-\beta ^{2}}\right) 
\left( 1+\frac{4\epsilon _{X}^{-}(1-\alpha ^{2})}
{\sqrt{\Delta _{\text{Z}}^{2}(1-\alpha ^{2})+\frac{\Delta _{\text{SAS}}^{2}\alpha ^{2}}
{1-\beta ^{2}}}}\right) =0.
\end{equation}
The solutions are 
\begin{equation}
\beta =\pm \sqrt{1-\left( \frac{\Delta _{\text{SAS}}}{\Delta _{\text{Z}}}
\right) ^{2}}+O(\Delta _{\text{SAS}}^{4}),  \label{limitsolution1} 
\end{equation}
with 
\begin{equation}
\Delta _{0}\rightarrow \Delta _{\text{SAS}}+O(\Delta _{\text{SAS}}^{3}),
\label{limitsolution2}
\end{equation}
for (\ref{EqD0}). By using (\ref{ImbalParam}) we have 
\begin{equation}
\mathcal{P}_{z}^{0}=\sigma _{0}=\pm \alpha ^{2}
+O(\Delta _{\text{SAS}}^{2}). 
\label{ImbalCanted}
\end{equation}
The parameters $\alpha $ and $\beta $ are simple functions of the physical
variables $\Delta _{\text{SAS}}/\Delta _{\text{Z}}$ and $\sigma _{0}$ in the
limit $\Delta _{\text{SAS}}\rightarrow 0$. 

In particular, one of the layers becomes empty in the pseudospin phase 
and also near the pseudospin-phase boundary in the CAF phase, 
since we have $\sigma _{0}\rightarrow \pm 1$ as $\alpha \rightarrow 1$. 
On the other hand, the bilayer system becomes balanced, 
since we have $\sigma _{0}\rightarrow 0$ as $\alpha \rightarrow 0$ in the spin phase
and also near the spin-phase boundary in the CAF phase. 
We might expect novel phenomena associated with the interlayer phase coherence
in the CAF phase.


\section{Effective Hamiltonian for Goldstone Modes}
\label{sec:III} 

Having reviewed the three phases in the bilayer system at $\nu =2$, we 
proceed to discuss the symmetry breaking pattern and construct the effective
Hamiltonian for the Goldstone modes in each phase. There is a systematic method for 
this purpose, 
which was developed in particle and nuclear physics\cite{Coleman:1969sm,Callan:1969sn}.

We analyze excitations around the classical ground state (\ref{orderparameter1}). 
It is convenient to introduce the SU(4) isospin notation
such that
\begin{equation}
\mathcal{I}_{a0}^{(0)}=\mathcal{S}_{a}^{0},\qquad \mathcal{I}_{0a}^{(0)}=
\mathcal{P}_{a}^{0},\qquad \mathcal{I}_{ab}^{(0)}=\mathcal{R}_{ab}^{0}.
\end{equation}
We set all of them into one 15-dimensional vector $\mathcal{I}_{\mu \nu
}^{(0)}$ with the index $\mu \nu $: Note that there is no component $\mathcal{I}_{00}^{(0)}$. 
Most general excitations are described by the
operator
\begin{equation}
\mathcal{I}_{\mu \nu }(x)=\mathcal{I}(x)\left[ \exp \left( i\sum_{\gamma
\delta }\pi _{\gamma \delta }{T}_{\gamma \delta }\right) \right] _{\mu \nu
}^{\mu ^{\prime }\nu ^{\prime }}\mathcal{I}_{\mu ^{\prime }\nu ^{\prime
}}^{0},  \label{su4isospin2}
\end{equation}
where ${T}_{\gamma \delta }$ are the matrices of the broken SU(4) generators
in the adjoint representation of SU(4), each of which is a $15\times 15$
matrix. The greek indices run over $0,x,y,z$. The phase field $\pi _{\gamma
\delta }(x)$ are the Goldstone modes associated with the broken generators,
and the coefficient $\mathcal{I}(x)$ is the amplitude function corresponding
to the ``sigma'' field in the linear sigma model.

It has been argued\cite{Ezawa:2005xi} that there are nine independent real
physical fields. They are the amplitude fluctuation field $\mathcal{I}(x)$
satisfying $\mathcal{I}^{2}(x)\leq 1$, and eight Goldstone modes $\pi
_{\gamma \delta }(x)$. Hence, only eight generating matrices ${T}_{\gamma
\delta }$ are involved in the formula (\ref{su4isospin2}). We shall
explicitly determine them in each phase in the following subsections. Since
we are only interested in an effective low energy theory of the Goldstone
bosons, we set $\mathcal{I}(x)=1$. Then we may identify 
\begin{equation}
\mathcal{S}_{a}=\mathcal{I}_{a0},\qquad \mathcal{P}_{a}=\mathcal{I}_{0a},
\qquad \mathcal{R}_{ab}=\mathcal{I}_{ab},  \label{IandSPR}
\end{equation}
and express various physical variables in terms of the Goldstone modes $\pi
_{\gamma \delta }(x)$.

We expand the formula \eqref{su4isospin2} in $\pi _{\gamma \delta }$, 
\begin{equation}
\mathcal{I}_{\mu \nu }(x)=\mathcal{I}_{\mu \nu }^{0}+\mathcal{I}_{\mu \nu
}^{(1)}(x)+\mathcal{I}_{\mu \nu }^{(2)}(x)+\cdots,  \label{expansion}
\end{equation}
where $\mathcal{I}_{\mu \nu }^{(n)}(x)$ is the $n$th order term in the
Goldstone mode $\pi _{\gamma \delta }$. Up to the second order, they are 
\begin{align}
\mathcal{I}_{\mu \nu }^{(1)}(x) &=-f_{\mu \nu ,\gamma \delta ,\mu ^{\prime
}\nu ^{\prime }}\pi _{\gamma \delta }\mathcal{I}_{\mu ^{\prime }\nu ^{\prime
}}^{0},  \label{expansion1} \\
\mathcal{I}_{\mu \nu }^{(2)}(x) &=\frac{1}{2!}f_{\mu \nu ,\gamma \delta
,\mu ^{\prime \prime }\nu ^{\prime \prime }}f_{\mu ^{\prime \prime }\nu
^{\prime \prime },\gamma ^{\prime }\delta ^{\prime },\mu^\prime \nu^\prime }\pi _{\gamma
\delta }\pi _{\gamma ^{\prime }\delta ^{\prime }}\mathcal{I}_{\mu^\prime \nu^\prime }^{0},
\label{expansion2}
\end{align}
where $f_{\mu \nu ,\gamma \delta ,\mu ^{\prime }\nu ^{\prime }}$ are the
structure constant of SU(4),
\begin{equation}
\left( T_{\gamma \delta }\right) _{\mu \nu }^{\mu ^{\prime }\nu ^{\prime 
}}=if_{\mu \nu ,\gamma \delta ,\mu ^{\prime }\nu ^{\prime }},
\end{equation}
about which we explain in Appendix A \eqref{su4commutator1}. 

Each phase is characterized by the order parameter $\mathcal{I}_{\mu \nu}^{0}$,
which are nothing but (\ref{orderparameter1}). 
The key observation is that 
 the first order term $\mathcal{I}_{\mu \nu }^{(1)}(x)$ contains all informations about 
 the symmetry breaking pattern and the associated
Goldstone modes, yielding their kinematic terms. On the other hand,
the second order term $\mathcal{I}_{\mu \nu }^{(2)}(x)$ provides them with gaps.

\subsection{Spin Phase}

First we analyze the spin phase. Setting $\alpha =0$ in the order parameters
(\ref{orderparameter1}), we obtain 
\begin{equation}
\mathcal{I}_{\mu \nu }^{0}=\delta _{\mu z}\delta _{\nu 0}.
\label{spinorderparameter}
\end{equation}
With the use of this, it is straighforward to calculate the first order term 
$\mathcal{I}_{\mu \nu }^{(1)}(x)$ in \eqref{expansion1}, 
\begin{equation}
\mathcal{I}_{x\mu }^{(1)}=-\pi _{y\mu },\ \mathcal{I}_{y\mu }^{(1)}=\pi
_{x\mu }.  \label{spinisospin}
\end{equation}
There are eight fields $\pi _{y\mu }$ and $\pi _{x\mu }$ with $\mu =0,x,y$
and $z$, which are the Goldstone modes. Since they emerge in eight
directions, $x\mu $ and $y\mu $, the broken generators are ${T}_{x\mu }$ and 
${T}_{y\mu }$. Consequently, the symmetry breaking pattern reads 
\begin{equation}
\text{SU}(4)\rightarrow \text{U}(1)\otimes \text{SU}(2)\otimes \text{SU}(2),  
\label{sssb}
\end{equation}
implying that the unbroken generators are ${T}_{z0}$, $T_{0a}$ and $T_{za}$.

We require \eqref{su4isospin2} to satisfy the SU(4) algebraic relation
\begin{equation}
\left[ \mathcal{I}_{a\mu }(\boldsymbol{x},t),\mathcal{I}_{b\mu }(\boldsymbol{y},t)\right] =i\epsilon _{abc}\rho _{\Phi }^{-1}\mathcal{I}_{c0}(\boldsymbol{x},t)\delta (\boldsymbol{x}-\mbox{\boldmath
$y$}),  \label{spinalgebra1}
\end{equation}
so that the field  $\mathcal{I}_{\mu\nu }$ describes the SU(4) isospin. 
From \eqref{spinalgebra1}, we
obtain the equal-time commutation relations for the Goldstone modes,  
\begin{equation}
\left[ \tilde{\pi}_{x\mu }(\boldsymbol{x},t),\tilde{\pi}_{y\mu }(\boldsymbol{y},t)\right] =i\delta (\boldsymbol{x}-\boldsymbol{y}),  \label{spinccr2}
\end{equation}
with $\tilde{\pi}_{\gamma \delta }=\rho _{\Phi }^{1/2}\pi _{\gamma \delta }$. 
Equivalently, we may construct a Lagrangian formalism so that (\ref{spinccr2}) is the canonical commutation relation.

It follows from (\ref{IandSPR}) and (\ref{spinisospin}) that the eight
Goldstone modes are explicitly given by
\begin{equation}
\mathcal{S}_{x}=-\pi _{y0},\quad \mathcal{S}_{y}=\pi _{x0},\quad \mathcal{R}
_{xa}=-\pi _{ya},\quad \mathcal{R}_{ya}=\pi _{xa}.
\label{SPRgoldstonespin}
\end{equation}
Substituting them into (\ref{su4effectivehamiltonian1}), we obtain the
effective Hamiltonian of the Goldstone modes in terms of the canonical sets of $\tilde{\pi}_{x\mu }$ and $\tilde{\pi}_{y\mu }$ as 
\begin{align}   
{\cal H}^{\text{spin}}&= \frac{2J_s}{\rho_0}\sum_{\mu=0,z}\left[(\partial_k\tilde{\pi}_{x\mu})^2+(\partial_k\tilde{\pi}_{y\mu})^2\right] \notag\\ 
&+ \frac{2J^d_s}{\rho_0}\sum_{a=x,y}\left[(\partial_k\tilde{\pi}_{xa})^2+(\partial_k\tilde{\pi}_{ya})^2\right] \notag\\ 
&-4\rho_\phi\epsilon_X^-{\cal I}^{(2)}_{z0}-2\epsilon_X^-\sum_{\mu=0,z}\left[(\tilde{\pi}_{x\mu})^2+(\tilde{\pi}_{y\mu})^2\right]\notag\\
&-\rho_\phi\left[ \Delta_{\text{Z}}{\cal I}^{(2)}_{z0}+\Delta_{\text{SAS}}{\cal I}^{(2)}_{0x}+\Delta_{\text{bias}}{\cal I}^{(2)}_{0z}\right],  
 \label{spineffectivehamiltonian1}
\end{align}
where $\mathcal{I}_{0a}^{(2)}$ are given by (\ref{expansion2}), and read
\begin{align}
\mathcal{I}_{z0}^{(2)}&=1-\frac{1}{2}\sum_{\mu=0,x,y,z}(\pi_{x\mu}^2+\pi_{y\mu}^2), \notag \\
\mathcal{I}_{0x}^{(2)} &=\frac{\pi_{xy}\pi_{yz}+\pi_{yz}\pi_{xy}-\pi_{yy}\pi_{xz}-\pi_{xz}\pi_{yy}}{2} ,  \notag \\
\mathcal{I}_{0z}^{(2)} &=\frac{\pi_{xx}\pi_{yy}+\pi_{yy}\pi_{xx}-\pi_{xy}\pi_{yx}-\pi_{yx}\pi_{xy}}{2}.
\end{align}
The annihilation operators are defined by
\begin{align}
\eta_1&=\frac{\tilde{\pi}_{x0}+i\tilde{\pi}_{y0}}{\sqrt{2}}, \quad
\eta_2=\frac{\tilde{\pi}_{xx}+i\tilde{\pi}_{yx}}{\sqrt{2}},\notag\\
\eta_3&=\frac{\tilde{\pi}_{xy}+i\tilde{\pi}_{yy}}{\sqrt{2}}, \quad 
\eta_4=\frac{\tilde{\pi}_{xz}+i\tilde{\pi}_{yz}}{\sqrt{2}}, 
\label{spincomplexfluctuation}
\end{align}
and satisfy the commutation relations,  
\begin{equation}
\left[ \eta_i(\boldsymbol{x},t),\eta_j^\dagger(\boldsymbol{y},t) \right]
=\delta_{ij}\delta(\boldsymbol{x}-\boldsymbol{y}),
\end{equation} 
with $i,j=1,2,3,4$. 

The effective Hamiltonian \eqref{spineffectivehamiltonian1} reads
in terms of the creation and annihilation variables \eqref{spincomplexfluctuation} as
\begin{align}
{\cal H}^{\text{spin}}&=\frac{4J_s}{\rho_0}\sum_{a=1,4}\partial_k \eta_a^\dagger\partial_k \eta_a
+\frac{4J^d_s}{\rho_0}\sum_{a=2,3}\partial_k \eta_a^\dagger\partial_k \eta_a\notag\\
&+\Delta_{\text{Z}}\sum_{a=1,4}\eta_a^\dagger\eta_a+[\Delta_{\text{Z}}+4\epsilon^-_X ]
\sum_{a=2,3}\eta_a^\dagger\eta_a\notag\\
&-\frac{\Delta_{\text{bias}}}{i}[\eta_2^\dagger\eta_3-\eta_3^\dagger\eta_2]
-\frac{\Delta_{\text{SAS}}}{i}[\eta_3^\dagger\eta_4-\eta_4^\dagger\eta_3].
\end{align} 
The variables $\eta _{2}$, $\eta _{3}$ and $\eta _{4}$ are mixing.

In the momentum space the annihilation and creation operators are 
$\eta_{i,\boldsymbol{k}}$ and $\eta_{i,\boldsymbol{k}}^\dagger$ together with the commutation relations, 
\begin{equation}
\left[ \eta_{i,\boldsymbol{k}},\eta_{j,\boldsymbol{k}^\prime}^\dagger\right]=\delta_{ij}\delta(\boldsymbol{k}-\boldsymbol{k}^\prime).
\end{equation}
For the sake of the simplicity
we consider the balanced configuration with $\Delta_{\text{bias}}=0$  
in the rest of this subsection.
Then the Hamiltonian density is given by
\begin{align}
H^{\text{spin}}&=\int d^2 k \ {\cal H}^{\text{spin}},\notag\\
{\cal H}^{\text{spin}}&={\cal H}^{\text{spin}}_1+{\cal H}^{\text{spin}}_2+{\cal H}^{\text{spin}}_3,
\label{totalspinhamiltonian}
\end{align} 
where 
\begin{align}
{\cal H}^{\text{spin}}_1&=\left[ \frac{4J_s}{\rho_0}\boldsymbol{k}^2+\Delta_{\text{Z}}\right]\eta_{1,\boldsymbol{k}}^\dagger\eta_{1,\boldsymbol{k}}, \label{momentumspinhamiltonian1} \\ 
{\cal H}^{\text{spin}}_2&=\left[ \frac{4J^d_s}{\rho_0}\boldsymbol{k}^2+\Delta_{\text{Z}}+4\epsilon^-_X \right]\eta_{2,\boldsymbol{k}}^\dagger\eta_{2,\boldsymbol{k}}, \label{momentumspinhamiltonian2} \\
{\cal H}^{\text{spin}}_3&=\left[ \frac{4J^d_s}{\rho_0}\boldsymbol{k}^2+\Delta_{\text{Z}}+4\epsilon^-_X\right]\eta_{3,\boldsymbol{k}}^\dagger\eta_{3,\boldsymbol{k}}
+\left[ \frac{4J_s}{\rho_0}\boldsymbol{k}^2+\Delta_{\text{Z}}\right]\notag\\
&\times \eta_{4,\boldsymbol{k}}^\dagger\eta_{4,\boldsymbol{k}} 
-\frac{\Delta_{\text{SAS}}}{i}\left[\eta_{3,\boldsymbol{k}}^\dagger\eta_{4,\boldsymbol{k}}-\eta_{4,\boldsymbol{k}}^\dagger\eta_{3,\boldsymbol{k}} \right].
\label{momentumspinhamiltonian3} 
\end{align}

We first analyze the dispersion relation and the coherence length of $ \eta_{1,\boldsymbol{k}}$. 
From \eqref{momentumspinhamiltonian1}, we have 
\begin{align}
E_{\eta_1}(\boldsymbol{k})&=\frac{4J_s}{\rho_0}\boldsymbol{k}^2+\Delta_{\text{Z}}, \label{spindispersion1} \\ 
\xi_{\eta_1}&=2l_B\sqrt{\frac{\pi J_s}{\Delta_{\text{Z}}}}. \label{coherencelength1}
\end{align}
The coherent length diverges in the limit $\Delta _{\text{Z}}\rightarrow 0$.
This mode is a pure spin wave since it describes the fluctuation of $\mathcal{S}_x$ and $\mathcal{S}_y$ as in \eqref{SPRgoldstonespin}. 
Indeed, the energy \eqref{spindispersion1} as well as the coherent length (\ref{coherencelength1})
depend only on the Zeeman gap $\Delta_{\text{Z}}$ and the intralayer stiffness $J_s$.   

We next analyze those of $\eta_{2,\boldsymbol{k}}$, 
\begin{align}
E_{\eta_2}(\boldsymbol{k})&=\frac{4J^d_s}{\rho_0}\boldsymbol{k}^2+\Delta_{\text{Z}}+4\epsilon^-_X,  \label{spindispersion2} \\
\xi_{\eta_2}&=2l_B\sqrt{\frac{\pi J^d_s}{\Delta_{\text{Z}}+4\epsilon^-_X}}. \label{coherencelength2}
\end{align} 
They depend not only $\Delta_{\text{Z}}$ but also on the exchange Coulomb energy $\epsilon_X^-$
and the interlayer stiffness originating in the interlayer Coulomb interaction. 

We finally analyze those of 
$\eta_{3,\boldsymbol{k}}$ and $\eta_{4,\boldsymbol{k}}$,
which are coupled. Hamiltonian \eqref{momentumspinhamiltonian3} can be written in the matrix form, 
\begin{align} 
{\cal H}^{\text{spin}}_3=\left( 
\begin{array}{c}
\eta_{3,\boldsymbol{k}}   \\
\eta_{4,\boldsymbol{k}}  \\
\end{array} 
\right)^\dagger
\left( 
\begin{array}{cc}
A_{\boldsymbol{k}} & -i\Delta_{\text{SAS}}  \\
i\Delta_{\text{SAS}} & B_{\boldsymbol{k}} \\
\end{array} 
\right)
\left( 
\begin{array}{c}
\eta_{3,\boldsymbol{k}}   \\
\eta_{4,\boldsymbol{k}}  \\
\end{array} 
\right),
\label{momentumspinmatrixhamiltonian}
\end{align} 
where
\begin{align}
A_{\boldsymbol{k}}=\frac{4J^d_s}{\rho_0}\boldsymbol{k}^2+\Delta_{\text{Z}}+4\epsilon^-_X, \quad
B_{\boldsymbol{k}}=\frac{4J_s}{\rho_0}\boldsymbol{k}^2+\Delta_{\text{Z}}.
\label{spinAandB}
\end{align}
Hamiltonian \eqref{momentumspinmatrixhamiltonian} can be diagonalized as 
\begin{align} 
{\cal H}^{\text{spin}}_3=\left( 
\begin{array}{c}
\tilde{\eta}_{3,\boldsymbol{k}}   \\
\tilde{\eta}_{4,\boldsymbol{k}}  \\
\end{array} 
\right)^\dagger
\left( 
\begin{array}{cc}
E^{\tilde{\eta}_3} & 0  \\
0 & E^{\tilde{\eta}_4} \\
\end{array} 
\right)
\left( 
\begin{array}{c}
\tilde{\eta}_{3,\boldsymbol{k}}   \\
\tilde{\eta}_{4,\boldsymbol{k}}  \\  
\end{array} 
\right),
\label{newspinmixinghamiltonian}
\end{align}
where 
\begin{align}
E^{\tilde{\eta}_3}&=\frac{1}{2}
\left[A_{\boldsymbol{k}}+B_{\boldsymbol{k}}+
\sqrt{(A_{\boldsymbol{k}}-B_{\boldsymbol{k}})^2+4\Delta^2_{\text{SAS}}} \right],\notag\\
E^{\tilde{\eta}_4}&=\frac{1}{2}
\left[A_{\boldsymbol{k}}+B_{\boldsymbol{k}}-
\sqrt{(A_{\boldsymbol{k}}-B_{\boldsymbol{k}})^2+4\Delta^2_{\text{SAS}}} \right],
\label{spindispersion3}
\end{align}
and the annihilation operator $\tilde{\eta}_{i,\boldsymbol{k}}$ ($i=3,4$) given by the form
\begin{align}
\tilde{\eta}_{3,\boldsymbol{k}}&=
\frac{-i\left(\sqrt{C^{2}_{\boldsymbol{k}}+4\Delta^2_{\text{SAS}}}+C_{\boldsymbol{k}}\right)\eta_{3,\boldsymbol{k}}
-2\Delta_{\text{SAS}}\eta_{4,\boldsymbol{k}}}
{\sqrt{2\left(C^{2}_{\boldsymbol{k}}+4\Delta^2_{\text{SAS}}+C_{\boldsymbol{k}} \sqrt{C^{2}_{\boldsymbol{k}}+4\Delta^2_{\text{SAS}}}
\right)}},\notag\\
\tilde{\eta}_{4,\boldsymbol{k}}&=
\frac{-i\left(\sqrt{C^{2}_{\boldsymbol{k}}+4\Delta^2_{\text{SAS}}}-C_{\boldsymbol{k}}\right)\eta_{3,\boldsymbol{k}}
+2\Delta_{\text{SAS}}\eta_{4,\boldsymbol{k}}}
{\sqrt{2\left(C^{2}_{\boldsymbol{k}}+4\Delta^2_{\text{SAS}}-C_{\boldsymbol{k}} \sqrt{C^{2}_{\boldsymbol{k}}+4\Delta^2_{\text{SAS}}}
\right)}},\notag\\
\label{newspinannihilation}
\end{align} 
with $C_{\boldsymbol{k}}=A_{\boldsymbol{k}}-B_{\boldsymbol{k}}$. 
The annihilation operators \eqref{newspinannihilation} 
satisfy the commutation relations
\begin{align}
\left[\tilde{\eta}_{i,\boldsymbol{k}},\tilde{\eta}^\dagger_{j,\boldsymbol{k}^\prime}\right]
=\delta_{ij}\delta(\boldsymbol{k}-\boldsymbol{k}^\prime), 
\end{align}
with $i,j=3,4$. 
We obtain the dispersions for the modes  $\tilde{\eta}_{i,\boldsymbol{k}}$ ($i=3,4$) 
from \eqref{spinAandB} and \eqref{spindispersion3}.

By taking the limit $\boldsymbol{k}\rightarrow0$ in \eqref{spindispersion3},
we have two gaps
\begin{align}
E^{\tilde{\eta}_3}_{\boldsymbol{k}=0}&=
\Delta_{\text{Z}}+2\epsilon^-_X+\left[4(\epsilon^-_X)^2+\Delta^2_{\text{SAS}}\right]^{\frac{1}{2}},\notag\\
E^{\tilde{\eta}_4}_{\boldsymbol{k}=0}&=
\Delta_{\text{Z}}+2\epsilon^-_X-\left[4(\epsilon^-_X)^2+\Delta^2_{\text{SAS}}\right]^{\frac{1}{2}}.
\label{spingaps}
\end{align}
The gapless condition $(E^{\tilde{\eta}_4}_{\boldsymbol{k}=0}=0)$ implies
\begin{equation}
\Delta_{\text{Z}}(\Delta_{\text{Z}}+4\epsilon^-_X)-\Delta_{\text{SAS}}^2=0,
\end{equation}
which holds only along the boundary of the spin and CAF phases: See (4.17) in Ref.\cite{Ezawa:2005xi}.
In the interior of the spin phase we have $\Delta_{\text{Z}}(\Delta_{\text{Z}}+4\epsilon^-_X)-\Delta_{\text{SAS}}^2>0$,
as implies that there arise no gapless modes from 
$\tilde{\eta}_3$ and $\tilde{\eta}_4$.
These excitation modes are residual spin waves coupled with the layer degree of freedom. 

\subsection{Pseudospin Phase}

We next analyze the pseudospin phase. Setting $\alpha =1$ in the order parameters (\ref{orderparameter1}), we obtain 
\begin{equation}
\mathcal{I}_{\mu \nu }^{0}=\sqrt{1-\beta ^{2}}\delta _{\mu 0}\delta _{\nu
x}+\beta \delta _{\mu 0}\delta _{\nu z}.  \label{ppinorderparameters2}
\end{equation}
In order to determine the symmetry breaking pattern, we rotate this vector
around the $0y$ axis so that only one component becomes nonzero. We can show 
that 
\begin{equation}
\mathcal{I}_{\mu \nu }^{\text{p}(0)}\equiv \left[ V_{\beta }(\theta
_{\beta })\right] _{\mu \nu }^{\mu ^{\prime }\nu ^{\prime }}\mathcal{I}_{\mu
^{\prime }\nu ^{\prime }}^{(0)}=\delta _{\mu 0}\delta _{\nu x}, 
\label{PpinOrder}
\end{equation} 
by choosing 
\begin{equation}
V_{\beta }(\theta _{\beta })=\exp (i\theta _{\beta }T_{0y}),
\label{ppinrotation}
\end{equation}
with $\cos \theta _{\beta }=\sqrt{1-\beta ^{2}}$ and $\sin \theta _{\beta
}=-\beta $.

In the rotated basis the order parameter has a single nonzero
component just as (\ref{spinorderparameter}) in the case of the spin phase.
Therefore the further analysis goes in parallel with that given in the
previous subsection. Namely, there are eight Goldstone fields,
\begin{equation}
\mathcal{I}_{\mu y}^{\text{p}(1)}=-\pi _{\mu z}^{\text{p}},\quad 
\mathcal{I}_{\mu z}^{\text{p}(1)}=\pi _{\mu y}^{\text{p}}, 
\label{ppinisospin}
\end{equation}
and the symmetry breaking pattern reads 
\begin{equation}
\text{SU}(4)\rightarrow\text{U}(1)\otimes \text{SU}(2)\otimes 
 \text{SU}(2),
\end{equation}
precisely as in the spin phase.

Let us relate the variables in the rotated system to the original variables
in the formula (\ref{su4isospin2}). The SU(4) isospin operator after the
rotation is given by
\begin{equation}
\mathcal{I}_{\mu \nu }^{\text{p}}(x)=\left[ V_{\beta }(\theta _{\beta })
\right] _{\mu \nu }^{\mu ^{\prime }\nu ^{\prime }}\mathcal{I}_{\mu ^{\prime
}\nu ^{\prime }}(x),  \label{OrigiToPpin}
\end{equation}
with the use of (\ref{ppinrotation}). We substitute (\ref{su4isospin2}) into
this formula to find
\begin{equation}
\mathcal{I}_{\mu \nu }^{\text{p}}(x)=\left[ \exp \left( i\sum_{\gamma \delta
}\pi _{\gamma \delta }^{\text{p}}{T}_{\gamma \delta }\right) \right] _{\mu
\nu }^{\mu ^{\prime }\nu ^{\prime }}\mathcal{I}_{\mu ^{\prime }\nu ^{\prime
}}^{\text{p}(0)},  \label{GmodePpin}
\end{equation}
with (\ref{PpinOrder}), 
where $\pi _{\gamma \delta }^{\text{p}}$ is defined by 
\begin{eqnarray}
\pi _{\gamma \delta }^{\text{p}} &=&\left[ V_{\beta }(\theta _{\beta })
\right] _{\gamma \delta }^{\gamma ^{\prime }\delta ^{\prime }}
\pi _{\gamma^{\prime }\delta ^{\prime }},  \label{ppingoldstone} \\
\mathcal{I}_{\gamma \delta }^{\text{p}(0)} &=&\left[ V_{\beta }(\theta
_{\beta })\right] _{\gamma \delta }^{\gamma ^{\prime }\delta ^{\prime }}
\mathcal{I}_{\gamma ^{\prime }\delta ^{\prime }}^{0},
\end{eqnarray}
while $\mathcal{I}_{\gamma \delta }^{\text{p}(0)}$ has been used by \eqref{PpinOrder}. 
Here, we have used the formula of the SU(N) group, 
\begin{align}
\sum_{b}T_{b}\Phi _{b}^{\prime }& =\sum_{b}T_{b}\left[ \text{exp}i\theta _{a}
\text{Ad}(T_{a})\right] _{b}^{c}\Phi _{c}  \notag \\ 
& =\text{exp}\left[ i\theta _{a}T_{a}\right] \Phi _{b}T_{b}\text{exp}\left[
-i\theta _{a}T_{a}\right] ,  \label{adjointtransformation}
\end{align}
where $\Phi _{b}$ is an arbitrary adjoint vector with $a,b,c=1,\ldots,\\
\text{dim\ SU(N)}$, and $\text{exp}\left[ i\theta _{a}T_{a}\right] $ is the
element of SU(N). Here we have N$=4$ and $\Phi _{b}$ corresponds to $\pi
_{\mu \nu }$. 

The SU(4) isospin density fields $\mathcal{I}_{\mu \nu}^{\text{p}}$ satisfy
the SU(4) algebraic relations 
\begin{equation}
\left[ \mathcal{I}_{\mu a}^{\text{p}}(\boldsymbol{x},t),
\mathcal{I}_{\mu b}^{\text{p}}(\boldsymbol{y},t)\right] 
=i\epsilon _{abc}\rho_{\Phi }^{-1}\mathcal{I}_{0c}^{\text{p}}(\boldsymbol{x},t)\delta (\boldsymbol{x}-\boldsymbol{y}),  \label{ppinalgebra1}
\end{equation}
from which we obtain the canonical commutation relations for the Goldstone modes, 
\begin{equation}
\left[ \tilde{\pi}_{\mu y}^{\text{p}}(\boldsymbol{x},t),
\tilde{\pi}_{\mu z}^{\text{p}}(\boldsymbol{y},t)\right] 
=i\delta (\mbox{\boldmath$x$}-\boldsymbol{y}),
\end{equation}
with $\tilde{\pi}_{\mu \nu }^{\text{p}}=\rho _{\Phi }^{1/2}
\pi _{\mu \nu }^{\text{p}}$.

We go on to derive the effective Hamiltonian governing these Goldstone
modes. The first step is to convert the relation (\ref{OrigiToPpin}) to
express the original fields in terms of those in the rotated system.
Explicitly we have
\begin{align}
\mathcal{I}_{\mu x}& =c_{\theta _{\beta }}\mathcal{I}_{\mu x}^{\text{p}}
+s_{\theta _{\beta }}\mathcal{I}_{\mu z}^{\text{p}},  \notag \\
\mathcal{I}_{\mu z}& =-s_{\theta _{\beta }}\mathcal{I}_{\mu x}^{\text{p}}
+c_{\theta _{\beta }}\mathcal{I}_{\mu z}^{\text{p}},  \notag \\
\mathcal{I}_{a0}& =\mathcal{I}_{a0}^{\text{p}},\quad \mathcal{I}_{y\mu }=
\mathcal{I}_{y\mu }^{\text{p}}.  \label{ppinisospinrelation}
\end{align}
The second step is to expand (\ref{GmodePpin}) in terms of $\pi _{\gamma
\delta }^{\text{p}}$, 
\begin{align}
\mathcal{I}_{\mu y}^{\text{p}}& =-\pi _{\mu z}^{\text{p}}+\mathcal{O}(\pi
^{2}),\ \mathcal{I}_{\mu z}^{\text{p}}=\pi _{\mu y}^{\text{p}}
+\mathcal{O}(\pi ^{2}), \notag \\
\mathcal{I}_{x0}^{\text{p}}& =\frac{\pi _{zz}^{\text{p}}\pi _{yy}^{\text{p}}+
\pi _{yy}^{\text{p}}\pi _{zz}^{\text{p}}-\pi _{zy}^{\text{p}}
\pi _{yz}^{\text{p}}-\pi _{yz}^{\text{p}}\pi _{zy}^{\text{p}}}{2}+\mathcal{O}(\pi ^{3}),
\notag \\
\mathcal{I}_{y0}^{\text{p}}& =\frac{\pi _{zy}^{\text{p}}\pi _{xz}^{\text{p}}
+\pi _{xz}^{\text{p}}\pi _{zy}^{\text{p}}-\pi _{zz}^{\text{p}}
\pi _{xy}^{\text{p}}-\pi _{xy}^{\text{p}}\pi _{zz}^{\text{p}}}{2}+\mathcal{O}(\pi ^{3}),
\notag \\
\mathcal{I}_{z0}^{\text{p}}& =\frac{\pi _{xy}^{\text{p}}\pi _{yz}^{\text{p}}
+\pi _{yz}^{\text{p}}\pi _{xy}^{\text{p}}-\pi _{yy}^{\text{p}}
\pi _{xz}^{\text{p}}-\pi _{xz}^{\text{p}}\pi _{yy}^{\text{p}}}{2}+\mathcal{O}(\pi ^{3}),
\notag \\
\mathcal{I}_{xx}^{\text{p}}& =-\frac{\pi _{xz}^{\text{p}}\pi _{0z}^{\text{p}}
+\pi _{0z}^{\text{p}}\pi _{xz}^{\text{p}}+\pi _{xy}^{\text{p}}
\pi _{0y}^{\text{p}}+\pi _{0y}^{\text{p}}\pi _{xy}^{\text{p}}}{2}+\mathcal{O}(\pi ^{3}),
\notag \\
\mathcal{I}_{yx}^{\text{p}}& =-\frac{\pi _{yz}^{\text{p}}\pi _{0z}^{\text{p}}
+\pi _{0z}^{\text{p}}\pi _{yz}^{\text{p}}+\pi _{yy}^{\text{p}}
\pi _{0y}^{\text{p}}+\pi _{0y}^{\text{p}}\pi _{yy}^{\text{p}}}{2}+\mathcal{O}(\pi ^{3}),
\notag \\
\mathcal{I}_{zx}^{\text{p}}& =-\frac{\pi _{zz}^{\text{p}}\pi _{0z}^{\text{p}}
+\pi _{0z}^{\text{p}}\pi _{zz}^{\text{p}}+\pi _{zy}^{\text{p}}
\pi _{0y}^{\text{p}}+\pi _{0y}^{\text{p}}\pi _{zy}^{\text{p}}}{2}+\mathcal{O}(\pi ^{3}),
\notag \\
\mathcal{I}_{0x}^{\text{p}}& =1-\sum_{\mu =0,x,y,z}
\frac{(\pi _{\mu y}^{\text{p}})^{2}+(\pi _{\mu z}^{\text{p}})^{2}}{2}+\mathcal{O}(\pi ^{3}).
\label{ppingoldstoneexpansion}
\end{align}
Now, using (\ref{IandSPR}) we obtain the expression of $\mathcal{S}_{a},
\mathcal{P}_{a},\mathcal{R}_{ab}$ in terms of $\pi _{\gamma \delta }^{\text{p}}$, which we substitute into the effective Hamiltonian (\ref{su4effectivehamiltonian1}).

In this way we derive the effective Hamiltonian of the Goldstone modes in
terms of the canonical sets of $\tilde{\pi}_{\mu y}^{\text{p}}$ and $\tilde{\pi}_{\mu z}^{\text{p}} $. In the momentum space it reads
\begin{align}
{\cal H}^{\text{p}}={\cal H}^{\text{p}}_1+{\cal H}^{\text{p}}_2+{\cal H}^{\text{p}}_3,
\label{totalppinhamiltonian}
\end{align}
where 
\begin{align}
{\cal H}^{\text{p}}_1&=C^{\text{p}}_{\boldsymbol{k}}\tilde{\pi}_{0y,\boldsymbol{k}}^{\text{p}\dagger}\tilde{\pi}^{\text{p}}_{0y,\boldsymbol{k}}
+B^{\text{p}}_{\boldsymbol{k}}\tilde{\pi}_{0z,\boldsymbol{k}}^{\text{p}\dagger}\tilde{\pi}^{\text{p}}_{0z,\boldsymbol{k}}, 
 \label{momentumppinhamiltonian1} \\   
{\cal H}^{\text{p}}_2&= 
A^{\text{p}}_{\boldsymbol{k}}\tilde{\pi}^{\text{p}\dagger}_{zy,\boldsymbol{k}}\tilde{\pi}^{\text{p}}_{zy,\boldsymbol{k}}
+B^{\text{p}}_{\boldsymbol{k}}\tilde{\pi}_{zz,\boldsymbol{k}}^{\text{p}\dagger}\tilde{\pi}^{\text{p}}_{zz,\boldsymbol{k}}, 
\label{momentumppinhamiltonian2} \\   
{\cal H}^{\text{p}}_3&=
(\vec{\tilde{\pi}}^{\text{p}})^\dagger 
\mathcal{M}^{\text{p}} 
\vec{\tilde{\pi}}^{\text{p}},
\label{momentumppinhamiltonian3}  
\end{align}
with 
\begin{align}
A^{\text{p}}_{\boldsymbol{k}}&=\frac{2J_1^\beta}{\rho_0}k^2+\frac{\Delta_{\text{SAS}}}{2\sqrt{1-\beta^2}}-2\epsilon^-_X(1-\beta^2),\notag\\
B^{\text{p}}_{\boldsymbol{k}}&=\frac{2J^d_s}{\rho_0}k^2+\frac{\Delta_{\text{SAS}}}{2\sqrt{1-\beta^2}},\notag\\
C^{\text{p}}_{\boldsymbol{k}}&=\frac{2J_1^\beta}{\rho_0}k^2+\frac{\Delta_{\text{SAS}}}{2\sqrt{1-\beta^2}}+\epsilon_{\text{cap}}(1-\beta^2),\notag\\
J_1^\beta&=(1-\beta^2)J_s+\beta^2 J^d_s,\notag\\
\vec{\tilde{\pi}}^{ \text{p}}&= 
(\tilde{\pi}^{\text{p}}_{yy,\boldsymbol{k}},\tilde{\pi}^{\text{p}}_{xz,\boldsymbol{k}}
,\tilde{\pi}^{\text{p}}_{xy,\boldsymbol{k}},\tilde{\pi}^{\text{p}}_{yz,\boldsymbol{k}}),
\notag\\
\mathcal{M}^{\text{p}} &= 
\left(  
\begin{array}{cccc} 
A^{\text{p}}_{\boldsymbol{k}} & \Delta_{\text{Z}}/2 & 0 & 0   \\
\Delta_{\text{Z}}/2 & B^{\text{p}}_{\boldsymbol{k}} & 0 & 0   \\ 
0 & 0 & A^{\text{p}}_{\boldsymbol{k}} & -\Delta_{\text{Z}}/2     \\
0 & 0 & -\Delta_{\text{Z}}/2 & B^{\text{p}}_{\boldsymbol{k}}     \\ 
\end{array} 
\right).
\label{ppinvariable}
\end{align}
The canonical commutation relations are 
\begin{align}
\left[ \tilde{\pi}^{\text{p}}_{\mu y,\boldsymbol{k}},\tilde{\pi}^{\text{p}}_{\mu z,\boldsymbol{k}^\prime}\right]&=i\delta(\boldsymbol{k}+\boldsymbol{k}^\prime), 
\label{momentumppinccr}
\end{align}   
for each $\mu=0,x,y,z$.

We first analyze the dispersions  and the coherence lengths of the canonical sets of 
the modes $\tilde{\pi}^{\text{p}}_{0y}$ and $\tilde{\pi}^{\text{p}}_{0z}$
from \eqref{momentumppinhamiltonian1}. 
Since  the ground state is a squeezed coherent state due to the capacitance energy $\epsilon_{\text{cap}}$, it is more convenient\cite{Ezawa:2008ae} to use the dispersion and the coherence lengths of $\tilde{\pi}^{\text{p}}_{0 y}$ and $\tilde{\pi}^{\text{p}}_{0 z}$ separately.
The dispersion relations are given by
\begin{align}
E_{\boldsymbol{k}}^{\tilde{\pi}^{\text{p}}_{0y}}&=\frac{2J^\beta_1}{\rho_0}\boldsymbol{k}^2+\frac{\Delta_{\text{SAS}}}{2\sqrt{1-\beta^2}}
+\epsilon_{\text{cap}}(1-\beta^2),\\
E_{\boldsymbol{k}}^{\tilde{\pi}^{\text{p}}_{0z}}&=\frac{2J_s^d}{\rho_0}\boldsymbol{k}^2+\frac{\Delta_{\text{SAS}}}{2\sqrt{1-\beta^2}},
\end{align}
and their coherence lengths are
\begin{align}
\xi^{\tilde{\pi}^{\text{p}}_{0y}}&=2l_B\sqrt{\frac{\pi J^\beta_1}{\frac{\Delta_{\text{SAS}}}{\sqrt{1-\beta^2}}
+ 2\epsilon_{\text{cap}}(1-\beta^2)}}, \\
\xi^{\tilde{\pi}^{\text{p}}_{0z}}&=2l_B\sqrt{\frac{\pi J^d_s\sqrt{1-\beta^2}}{\Delta_{\text{SAS}}}}.  
\end{align}
They describe a pseudospin wave.
 
The similar analysis can be adopted for the canonical sets of $\tilde{\pi}^{\text{p}}_{zy}$ and $\tilde{\pi}^{\text{p}}_{zz}$ in \eqref{momentumppinhamiltonian2}. 
The dispersion relations are given by 
\begin{align}
E_{\boldsymbol{k}}^{\tilde{\pi}^{\text{p}}_{zy}}&=\frac{2J^\beta_1}{\rho_0}\boldsymbol{k}^2+\frac{\Delta_{\text{SAS}}}{2\sqrt{1-\beta^2}}
-2\epsilon^-_X(1-\beta^2),\label{StepA}\\
E_{\boldsymbol{k}}^{\tilde{\pi}^{\text{p}}_{zz}}&=\frac{2J_s^d}{\rho_0}\boldsymbol{k}^2+\frac{\Delta_{\text{SAS}}}{2\sqrt{1-\beta^2}}.
\end{align}
Their coherence lengths are
\begin{align}
\xi^{\tilde{\pi}^{\text{p}}_{zy}}&=2l_B\sqrt{\frac{\pi J^\beta_1}{\frac{\Delta_{\text{SAS}}}{\sqrt{1-\beta^2}}
-4\epsilon_X^-(1-\beta^2)}}, \\  
\xi^{\tilde{\pi}^{\text{p}}_{zz}}&=2l_B\sqrt{\frac{\pi J^d_s\sqrt{1-\beta^2}}{\Delta_{\text{SAS}}}}.  
\end{align}  
It appears that $\xi ^{\tilde{\pi}_{zy}^{\text{p}}}$ is ill-defined for $%
\Delta _{\text{SAS}}\rightarrow 0$ in (\ref{StepA}). This is not the case
due to the relation (\ref{StepB}) in the pseudospin phase, which we mention
soon.

Finally,  making an analysis of  the Hamiltonian \eqref{momentumppinhamiltonian3}
as in the case of the spin phase, 
we obtain the condition for the  
existence of a gapless mode,
 \begin{equation}
\frac{\Delta_{\text{SAS}}}{\sqrt{1-\beta^2}}
\left[ 
\frac{\Delta_{\text{SAS}}}{\sqrt{1-\beta^2}}-4\epsilon^-_X(1-\beta^2)
\right]-\Delta^2_{\text{Z}}=0. 
\end{equation}
It occurs along the pseudospin-canted boundary: See (5.3) and (5.4) in Ref.\cite{Ezawa:2005xi}. 
Inside the pseudospin phase, since we have
\begin{equation}
\frac{\Delta_{\text{SAS}}}{\sqrt{1-\beta^2}}
\left[ 
\frac{\Delta_{\text{SAS}}}{\sqrt{1-\beta^2}}-4\epsilon^-_X(1-\beta^2)  
\right]-\Delta^2_{\text{Z}}>0, \label{StepB}
\end{equation}
there are no gapless modes.

\subsection{CAF phase} 

Finally we analyze the CAF phase. This phase is
characterized by the order parameters \eqref{orderparameter1}, which we may
rewrite as 
\begin{align}
\mathcal{I}_{\mu \nu }^{(0)} &=c_{\theta_\delta}c_{\theta_\alpha}\delta _{\mu
z}\delta _{\nu 0}+s_{\theta_\delta}s_{\theta_\alpha}\left( c_{\theta _{\beta
}}\delta _{\mu 0}\delta _{\nu x}-s_{\theta _{\beta }}\delta _{\mu 0}\delta
_{\nu z}\right)   \notag \\
&+s_{\theta_\delta}c_{\theta_\alpha}s_{\theta _{\beta }}\delta_{\mu x}\delta _{\nu x}  
-c_{\theta_\delta}s_{\theta_\alpha}\delta _{\mu y}\delta _{\nu y}   
+s_{\theta_\delta}c_{\theta_\alpha}c_{\theta _{\beta }}\delta
_{\mu x}\delta _{\nu z},
\label{caforderparameters0}  
\end{align}
where 
\begin{align}
c_{\theta_\alpha} &\equiv \cos \theta_\alpha=\sqrt{1-\alpha^2}, \quad
s_{\theta_\alpha} \equiv \sin {\theta_\alpha}=\alpha, \notag\\
c_{\theta_\beta} &\equiv \cos {\theta_\beta}=\sqrt{1-\beta^2}, \quad
s_{\theta_\beta} \equiv \sin {\theta_\beta}=-\beta, \notag\\
c_{\theta_\delta}&\equiv \cos \theta_\delta=\frac{\Delta_{Z}\sqrt{1-\beta^2}}{\Delta_0}\sqrt{1-\alpha^2}, \quad
s_{\theta_\delta} \equiv \sin \theta_\delta=\frac{\Delta_{\text{SAS}}}{\Delta_0}\alpha.
\label{sincosdefinition} 
\end{align}
   
The order parameter  $\mathcal{I}_{\mu \nu }^{(0)}$ is quite
complicated. Nevertheless, the problem is just to find an appropriate
rotation in the SU(4) space so that the order parameter has only a single
nonzero component after the rotation. 

There are two ways.  One is by choosing the rotational transformation as
\begin{equation}
U^s_{\alpha,\beta}=\text{exp}[i\theta_\delta T_{yz}]\text{exp}[i\theta_\alpha T_{xy}]V_\beta(\theta_\beta),   
\label{SCAFtransformation} 
\end{equation}
with $V_\beta$ given by \eqref{ppinrotation}, and we obtain   
\begin{equation}
\mathcal{I}_{\mu \nu }^{\text{sc}(0)}\equiv \left[U^s_{\alpha,\beta} \right] _{\mu \nu
}^{\mu ^{\prime }\nu ^{\prime }}\mathcal{I}_{\mu ^{\prime }\nu ^{\prime
}}^{(0)}=\delta _{\mu z}\delta _{\nu 0}. 
\end{equation}
In this rotated basis, the further analysis goes in parallel with that given
in the spin phase. Another choice of the rotational transformation is given by 
\begin{align}
U^p_{\alpha,\beta}&=\text{exp}\left[i\left(\theta_\delta-\frac{\pi}{2}\right) T_{yz}\right]
\text{exp}\left[i\left(\theta_\alpha-\frac{\pi}{2}\right) T_{xy}\right]
 V_\beta(\theta_\beta),\notag\\    
&=\text{exp}\left[-i\frac{\pi}{2} T_{yz}\right] 
\text{exp}\left[-i\frac{\pi}{2}T_{xy}\right]U^s_{\alpha,\beta},
\label{PCAFtransformation}  
\end{align} 
obtaining  
\begin{equation}
\mathcal{I}_{\mu \nu }^{\text{pc}(0)}\equiv \left[U^p_{\alpha,\beta} \right] _{\mu \nu
}^{\mu ^{\prime }\nu ^{\prime }}\mathcal{I}_{\mu ^{\prime }\nu ^{\prime
}}^{(0)}=\delta _{\mu 0}\delta _{\nu x}. 
\end{equation} 
In this rotated basis, the further analysis goes in parallel with that given
in the pseudospin phase. We call the rotated basis of the SU(4) group given by \eqref{SCAFtransformation}, the s-coordinate, and  
the rotated basis given by \eqref{PCAFtransformation}, the p-coordinate. 
They give the identical results.

We  make an analysis by employing the s-coordinate.  
Namely, we define the SU(4) isospin operator in
the s-coordinate by  
\begin{align}
\mathcal{I}_{\mu \nu }^{\text{sc}}(x)&=\left[ U^s_{\alpha,\beta}\right] _{\mu \nu }^{\mu
^{\prime }\nu ^{\prime }}\mathcal{I}_{\mu ^{\prime }\nu ^{\prime }}(x) \notag\\ 
&=\left[ \exp \left( i\sum_{\gamma
\delta }\pi^{\text{sc}} _{\gamma \delta }T_{\gamma \delta }\right) \right] _{\mu \nu
}^{\mu ^{\prime }\nu ^{\prime }}\mathcal{I}_{\mu ^{\prime }\nu ^{\prime 
}}^{\text{sc}(0)},   
\nonumber\\  
\label{SCAFbasis}  
\end{align}   
where 
\begin{equation}
\pi _{\gamma \delta }^{\text{sc}}=\left[ U^s_{\alpha,\beta}\right] _{\gamma \delta }^{\gamma ^{\prime } \delta^{\prime }}\pi
_{\gamma ^{\prime } \delta^{\prime } }
\label{SCAFgoldstone}
\end{equation} 
with \eqref{su4isospin2} and \eqref{adjointtransformation}. 

The eight Goldstone fields are,
\begin{equation}
\mathcal{I}_{x\mu }^{\text{sc}(1)}=-\pi^{\text{sc}} _{y\mu},\quad   
\mathcal{I}_{y\mu }^{\text{sc}(1)}=\pi^{\text{sc}} _{x\mu },
\label{SCAFgoldstone1} 
\end{equation}
 and the symmetry breaking pattern reads  
\begin{equation}
\text{SU}(4)\rightarrow \text{U}(1)\otimes\text{SU}(2)\otimes 
 \text{SU}(2),
\end{equation}
just as in the cases of the spin/pseudospin phase.  

Here we remark how the Goldstone modes in the CAF phase are transformed into those in spin/pseudospin phase at the phase boundary. 
On one hand, 
the field $\pi^{\text{sc}} _{\mu\nu}$ shift smoothly to the field \eqref{SPRgoldstonespin},
by the inverse transformation of \eqref{SCAFtransformation}, or 
by taking  the limit $\alpha,\beta \rightarrow0$,  as 
\begin{equation}
\pi^{\text{sc}} _{\mu\nu}\rightarrow\pi _{\mu\nu},  \label{sccoresppondence1}
\end{equation}
so that subscript of  $\pi^{\text{sc}} _{\mu\nu}$ perfectly matches with $\pi _{\mu\nu}$ for each $\mu\nu$ in the spin phase. 
On the other hand, $\pi^{\text{sc}} _{\mu\nu}$ shift smoothly to \eqref{ppingoldstone},
by  the inverse transformation of \\
$\text{exp}(i\theta_\delta T_{yz})
\text{exp}(i\theta_\alpha T_{xy})$,
or  taking the limit $\alpha\rightarrow1$ as 
\begin{align}
\pi^{\text{sc}} _{x0}&\rightarrow -\pi^{\text{p}} _{zz},\quad  
\pi^{\text{sc}} _{y0}\rightarrow\pi^{\text{p}} _{zy}, \notag\\
\pi^{\text{sc}} _{xx}&\rightarrow -\pi^{\text{p}} _{0z},\quad  
\pi^{\text{sc}} _{yx}\rightarrow\pi^{\text{p}} _{0y},\notag\\
\pi^{\text{sc}} _{xz}&\rightarrow\pi^{\text{p}} _{yy},\quad 
\pi^{\text{sc}} _{yz}\rightarrow\pi^{\text{p}} _{yz},  \notag\\
\pi^{\text{sc}} _{xy}&\rightarrow\pi^{\text{p}} _{xy},\quad   
\pi^{\text{sc}} _{yy}\rightarrow\pi^{\text{p}} _{xz}, \label{sccoresppondence2}
\end{align} 
for the fields in the pseudospin phase.

We require \eqref{SCAFbasis} to satisfy the SU(4) algebraic relation,    
\begin{equation} 
\left[{\cal I}^{\text{sc}}_{x\mu} (\boldsymbol{x},t),{\cal I}^{\text{sc}}_{y\mu} (\boldsymbol{y},t)\right]=
i\rho^{-1}_{\Phi}{\cal I}^{\text{sc}}_{z0} (\boldsymbol{x},t)\delta(\boldsymbol{x}-\boldsymbol{y})\label{scafalgebra1},
\end{equation}
from which we obtain the canonical commutation relation,
\begin{align} 
\left[\tilde{\pi}^{\text{sc}}_{x\mu} (\boldsymbol{x},t),\tilde{\pi}^{\text{sc}}_{y\mu} (\boldsymbol{y},t)\right]
=i\delta(\boldsymbol{x}-\boldsymbol{y}),
\label{scantedcanonicalconjugates} 
\end{align} 
with $\tilde{\pi}^{\text{sc}}_{\mu\nu}=\rho_\Phi^{1/2}\pi^{\text{sc}}_{\mu\nu}$.  

We are able to derive the effective Hamiltonian for the Goldstone modes precisely as we did for the pseudospin phase. Namely, we obtain the relations between the original fields 
$\mathcal{I}_{\mu \nu }$ and the fields $\pi _{\gamma \delta }^{\text{sc}}$
from (\ref{SCAFbasis}). We give the explicit relations in Appendix: See 
\eqref{cafisospinrelation}, and \eqref{cafgoldstoneexpansion}.  
Thus we derive the effective Hamiltonian of the Goldstone modes in terms of the canonical sets of
$\tilde{\pi}^{\text{sc}}_{x\mu }$ and $\tilde{\pi}_{y\mu }^{\text{sc}}$. 
Working in the momentum space,  the effective Hamiltonian reads, 
\begin{align}
{\cal{H}}^{\text{sc}}={\cal{H}}^{\text{sc}}_{1}+{\cal{H}}^{\text{sc}}_{2}, 
\end{align}
where
\begin{align}  
{\cal{H}}^{\text{sc}}_{1}&=      
G^{\text{c}}_{1,\boldsymbol{k}}(\tilde{\pi}^{\text{sc}}_{x0,\boldsymbol{k}})^{\dagger} \tilde{\pi}^{\text{sc}}_{x0,\boldsymbol{k}}+
G^{\text{c}}_{2,\boldsymbol{k}}(\tilde{\pi}^{\text{sc}}_{y0,\boldsymbol{k}})^{\dagger} \tilde{\pi}^{\text{sc}}_{y0,\boldsymbol{k}},
\label{scantedhamiltonian1}\\
{\cal{H}}^{\text{sc}}_{2}&=\vec{\pi}^{\text{sc}\dagger}_{\boldsymbol{k}} 
{\cal{M}}^{\text{sc}}_2\vec{\pi}^{\text{sc}}_{\boldsymbol{k}},
\label{scantedhamiltonian2}
\end{align} 
with   
\begin{align}
G^{\text{c}}_{1,\boldsymbol{k}}&=\frac{2}{\rho_0} J^{{\alpha}}_1 \boldsymbol{k}^2+\frac{\Delta_0 c^{-1}_{\theta_{\beta}}}{2}, \notag\\ 
G^{\text{c}}_{2,\boldsymbol{k}}&=\frac{2}{\rho_0}(c_{\theta_\delta}^2 J_s+s_{\theta_\delta}^2 J_1^\beta) \boldsymbol{k}^2+\frac{M-4 (s^2_{\theta_\delta} c_{\theta_\beta}^2
 +c_{\theta_\delta}^2)\epsilon^{-}_X}{2},\notag\\
J^{{\alpha}}_1&=c_{\theta_\alpha}^2 J_s+s_{\theta_\alpha}^2 J^d_s, \quad
M=4c_{\theta_\alpha}^2\epsilon^-_X 
+\Delta_0 c_{\theta_\beta}^{-1},
\label{cafcoefficients1}
\end{align}  
and
\begin{align}
\vec{\pi}^{\text{sc}}_{\boldsymbol{k}}=  
\left(
\begin{array}{c}
\tilde{\pi}^{\text{sc}}_{xx,\boldsymbol{k}}   \\
\tilde{\pi}^{\text{sc}}_{xz,\boldsymbol{k}}  \\
\tilde{\pi}^{\text{sc}}_{yy,\boldsymbol{k}}   \\ 
\tilde{\pi}^{\text{sc}}_{yx,\boldsymbol{k}}   \\
\tilde{\pi}^{\text{sc}}_{yz,\boldsymbol{k}}   \\
\tilde{\pi}^{\text{sc}}_{xy,\boldsymbol{k}}   \\
\end{array}
\right), \quad
{\cal{M}}_2^{\text{sc}}=  
\left(
\begin{array}{cccccc}
A^{\text{c}} &  c^{\text{c}}  & e^{\text{c}} & 0 & 0 & 0   \\
c^{\text{c}} & C^{\text{c}} & f^{\text{c}} & 0 & 0 & 0 \\
e^{\text{c}}  & f^{\text{c}}  & F^{\text{c}} & 0 & 0 & 0  \\ 
0 & 0 & 0 & B^{\text{c}} &  a^{\text{c}}  & b^{\text{c}}  \\
0 & 0 & 0 & a^{\text{c}} & D^{\text{c}} & d^{\text{c}}  \\
0 & 0 & 0 & b^{\text{c}}  & d^{\text{c}}  & E^{\text{c}}  \\
\end{array}
\right).
\label{matrixscaf}
\end{align}
The Matrix elements in \eqref{matrixscaf}  are given by 
\begin{align} 
A^{\text{c}}&=\frac{2\boldsymbol{k}^2}{\rho_0}\left[c_{\theta_\delta}^2 J_3^\beta+s_{\theta_\delta}^2 J^d_s\right]
+\frac{M}{2}-2 s_{\theta_\beta}^2 c_{\theta_\delta}^2\epsilon_X^-, \nonumber\\ 
B^{\text{c}}&=\frac{2\boldsymbol{k}^2}{\rho_0}\left[c_{\theta_\alpha}^2 J_3^\beta +s_{\theta_\alpha}^2 J_1^\beta\right]+\frac{\Delta_0 }{2c_{\theta_\beta}}
+\frac{c_{\theta_\beta}^2\epsilon_\alpha }{2}, \notag\\
C^{\text{c}}&=\frac{2\boldsymbol{k}^2}{\rho_0} J^\beta_1 +\frac{M}{2}-2 c_{\theta_\beta}^2\epsilon_X^-,\nonumber\\ 
D^{\text{c}}&=\frac{2\boldsymbol{k}^2}{\rho_0}\left[c_{\theta_\delta}^2 \left(s_{\theta_\alpha}^2 J_3^\beta+c_{\theta_\alpha}^2 J_1^\beta\right)+s_{\theta_\delta}^2 J^{\alpha}_1\right]
+\frac{\Delta_0 }{2c_{\theta_\beta}}
+
\frac{c_{\theta_\delta}^2 s_{\theta_\beta}^2\epsilon_\alpha }{2}, \notag\\
E^{\text{c}}&=\frac{2\boldsymbol{k}^2}{\rho_0}\left[s_{\theta_\delta}^2 \left(c_{\theta_\alpha}^2 J^\beta_3+s_{\theta_\alpha}^2 J^\beta_1\right)+c_{\theta_\delta}^2J^\alpha_3 \right] 
+\frac{M}{2}\notag\\
&+s_{\theta_\beta}^2 s_{\theta_\delta}^2 c_{\theta_\alpha}^2\epsilon_{\text{cap}}
-2 (c_{\theta_\beta}^2 s_{\theta_\delta}^2+c_{\theta_\delta}^2) s_{\theta_\alpha}^2\epsilon_X^-, \nonumber\\
F^{\text{c}}&=\frac{2\boldsymbol{k}^2}{\rho_0} J^d_s +\frac{M}{2},
\end{align}
and   
\begin{align}
a^{\text{c}}&=\frac{2\boldsymbol{k}^2}{\rho_0}c_{\theta_\delta}c_{2\theta_\alpha} J_2^{\beta}+\frac{s_{2\theta_\beta}  c_{\theta_\delta}}{4} \epsilon_{\alpha}, \notag\\ 
b^{\text{c}}&=-\frac{2\boldsymbol{k}^2}{\rho_0}s_{\theta_\delta} s_{2\theta_\alpha}  J_2^\beta +L+\frac{\Delta_{\text{SAS}}}{4\Delta_0}c_{\theta_\alpha} s_{2\theta_\beta} \epsilon_{\alpha},\notag\\ 
c^{\text{c}}&=\frac{2\boldsymbol{k}^2}{\rho_0}c_{\theta_\delta} J_2^\beta  + s_{2\theta_\beta}  c_{\theta_\delta}\epsilon_X^-, \notag\\ 
d^{\text{c}}&=-\frac{s_{2\theta_\alpha} s_{2\theta_\delta}}{4}[\frac{2\boldsymbol{k}^2}{\rho_0} \left(J_1^\beta+J^d_s-J_3^\beta-J_s\right) \notag\\
&+ s_{\theta_\beta}^2(2\epsilon_X^- -\epsilon_{\text{cap}})]-\frac{N}{2},\notag\\ 
e^{\text{c}}&=-\frac{L}{2}, \quad f^{\text{c}}=\frac{N}{2},    
\end{align} 
with  
\begin{align} 
J_1^\beta&=c_{\theta_\beta}^2 J_s+s_{\theta_\beta}^2 J^d_s, \  
J_2^\beta=\frac{s_{2\theta_\beta}}{2}  (J^d_s-J_s), \nonumber\\ 
J_3^\beta&=c_{\theta_\beta}^2 J^d_s+s_{\theta_\beta}^2 J_s, \ J_3^{\alpha}=c_{\theta_\alpha}^2 J^d_s+s_{\theta_\alpha}^2 J_s,
\nonumber\\ 
L&=-\frac{s_{2\theta_\beta}}{2} \left[s_{\theta_\delta} s_{2\theta_\alpha} (2\epsilon_X^- -\epsilon_{\text{cap}})
+c_{\theta_\alpha}\frac{\Delta_{\text{SAS}}}{\Delta_0}\epsilon_{\alpha}\right],\nonumber\\
N&=\frac{s_{2\theta_\delta} s_{2\theta_\alpha}   s_{\theta_\beta}^2}{2} 
 (2\epsilon_X^- -\epsilon_{\text{cap}}) 
+\frac{\Delta_{\text{SAS}}}{\Delta_0}(c_{\theta_\delta}c_{\theta_\alpha} s_{\theta_\beta}^2 \epsilon_{\alpha}+\Delta_Z),\notag\\ 
\epsilon_{\alpha}&=4c^2_{\theta_\alpha} \epsilon _X^-+2s_{\theta_\alpha}^2\epsilon_{\text{cap}},
\end{align}   
where we denote $s_{2\theta_{\alpha}}=\sin{2\theta_{\alpha}}$, $s_{2\theta_{\beta}}=\sin {2\theta_{\beta}}$, 
and $s_{2\theta_{\delta}}=\sin{2\theta_{\delta}}$.

It can be verified that the effective Hamiltonian \eqref{scantedhamiltonian1} and \eqref{scantedhamiltonian2}
reproduce the effective Hamiltonian in the spin phase \eqref{totalspinhamiltonian}, 
by taking the limit $\alpha\rightarrow0$ first,  and then diagonalize this Hamiltonian with 
the transformation  $V_\beta^{-1}$, 
or taking $\alpha,\beta\rightarrow0$.   
On the other hand, we reproduce the effective Hamiltonian in the pseudospin phase \eqref{totalppinhamiltonian},
by taking the limit $\alpha\rightarrow1$, in \eqref{scantedhamiltonian1} and \eqref{scantedhamiltonian2}.       

\subsection{CAF phase in $\Delta_{\text{SAS}}\rightarrow0$} 

The effective Hamiltonian in the CAF phase is too complicated to make a further analysis. 
We take the limit  $\Delta_{\text{SAS}}\rightarrow0$ to examine 
if some simplified formulas are obtained. 
In particular we would like to seek for gapless modes. 
Such gapless modes will play an important role to drive
the interlayer coherence in the CAF phase.  

In this limit we have 
\begin{align}
c_{\theta_\beta}&=\frac{\Delta_{\text{SAS}}}{\Delta_{\text{Z}}}, \quad 
s_{\theta_\beta}=\pm \sqrt{1-\left(\frac{\Delta_{\text{SAS}}}{\Delta_{\text{Z}}}\right)^2} \notag\\
c_{\theta_\delta}&=c_{\theta_\alpha}, \quad s_{\theta_\delta}=s_{\theta_\alpha}, \quad \Delta_0c^{-1}_{\theta_\beta}=\Delta_{\text{Z}},\notag\\
a^{\text{c}}&=b^{\text{c}}=c^{\text{c}}=e^{\text{c}}=L=0.
\end{align}
By using the above equations,  \eqref{scantedhamiltonian1} become 
\begin{align}
{\cal{H}}^{\text{sc}}_{1}=\left[\frac{4}{\rho_0}J^\alpha_1 k^2+\Delta_{\text{Z}} \right]
\eta^{\text{sc}\dagger}_{1,\boldsymbol{k}}\eta^{\text{sc}}_{1,\boldsymbol{k}},
\label{limitscantedhamiltonian1}
\end{align}
with
\begin{align}
\eta^{\text{sc}}_{1,\boldsymbol{k}}=\frac{\tilde{\pi}^{\text{sc}}_{x0,\boldsymbol{k}}+
i\tilde{\pi}^{\text{sc}}_{y0,\boldsymbol{k}}}{\sqrt{2}}.
\label{cafcoherentmode1} 
\end{align}
From \eqref{limitscantedhamiltonian1} we have the dispersion and the coherence length for mode $\eta^\text{sc}_1$ 
\begin{align}
E^{\eta^\text{sc}_1}&=\frac{4}{\rho_0}J^\alpha_1 \boldsymbol{k}^2+\Delta_{\text{Z}},  \quad
\ \xi^{\eta^\text{sc}_1}=2l_B\sqrt{\frac{\pi J_1^{{\alpha}} }{\Delta_{\text{Z}} }}.  
\label{dispersioncoherenceeta1s}
\end{align}
This mode is reminiscent of the spin wave ($\ref{spindispersion1}$) in the spin phase.  

We next investigate ${\cal{H}}^{\text{sc}}_{2}$ in \eqref{scantedhamiltonian2}. 
It yields
\begin{align}
{\cal{H}}^{\text{sc}}_{2}&={\cal{H}}^{\text{sc}}_{2,1}+{\cal{H}}^{\text{sc}}_{2,2},
\label{limitscantedhamiltonian2}\\
{\cal{H}}^{\text{sc}}_{2,1}&=\left[\frac{4}{\rho_0}J^\alpha_1 \boldsymbol{k}^2+\Delta_{\text{Z}} \right]
\eta^{\text{sc}\dagger}_{2,\boldsymbol{k}}\eta^{\text{sc}}_{2,\boldsymbol{k}},\label{limitscantedhamiltonian2-1}\\
{\cal{H}}^{\text{sc}}_{2,2}&=
\vec{\pi}^{\text{sc}\dagger}_{2,\boldsymbol{k}} 
{\cal{M}}^{\text{sc}}_{2,2}\vec{\pi}^{\text{sc}}_{2,\boldsymbol{k}},
\label{limitscantedhamiltonian2-2}
\end{align}
where 
\begin{align}
\eta^{\text{sc}}_{2,\boldsymbol{k} 
}=\frac{\tilde{\pi}^{\text{sc}}_{xx,\boldsymbol{k}}+
i\tilde{\pi}^{\text{sc}}_{yx,\boldsymbol{k}}}{\sqrt{2}}, 
\label{cafcoherentmode2} 
\end{align} 
and
\begin{align}
\vec{\pi}^{\text{sc}}_{2,\boldsymbol{k}}=\left( 
\begin{array}{c}
\tilde{\pi}^{\text{sc}}_{xz,\boldsymbol{k}} \\
\tilde{\pi}^{\text{sc}}_{yy,\boldsymbol{k}}  \\
\tilde{\pi}^{\text{sc}}_{yz,\boldsymbol{k}}      \\
\tilde{\pi}^{\text{sc}}_{xy,\boldsymbol{k}}  \\
\end{array}
\right), \quad
\mathcal{M}^{\text{sc}}_{2,2}= 
\left(
\begin{array}{cccc}
\tilde{C}^{\text{c}} &  \tilde{f}^{\text{c}} & 0 & 0 \\
\tilde{f}^{\text{c}} &\tilde{F}^{\text{c}} & 0 & 0 \\
0 & 0 & \tilde{D}^{\text{c}} &  \tilde{d}^{\text{c}}     \\
0 & 0 & \tilde{d}^{\text{c}} &\tilde{D}^{\text{c}}  \\
\end{array}
\right),
\label{limitedmatrixscaf}
\end{align}
with  
\begin{align}
\tilde{C}^{\text{c}}&=\frac{2\boldsymbol{k}^2}{\rho_0}J^d_s +\frac{\Delta_{\text{Z}}}{2}+2\epsilon^-_X \left(c^2_{\theta_{\alpha}}- \frac{\Delta_{\text{SAS}}^2}{\Delta_{\text{Z}}^2}\right) ,\notag\\
\tilde{F}^{\text{c}}&=\frac{2\boldsymbol{k}^2}{\rho_0}J^d_s 
+\frac{\Delta_{\text{Z}}}{2}+2c^2_{\theta_{\alpha}}\epsilon^-_X  ,\notag\\
\tilde{D}^{\text{c}}&=\frac{2\boldsymbol{k}^2}{\rho_0}(c^2_{\theta_{\alpha}}J^\alpha_3+s^2_{\theta_{\alpha}}J^\alpha_1 )
+\frac{\Delta_{\text{Z}}}{2}+ \frac{c^2_{\theta_{\alpha}}\epsilon_\alpha}{2},\notag\\ 
\tilde{d}^{\text{c}}&=\frac{\boldsymbol{k}^2}{\rho_0}s^2_{2\theta_{\alpha}}(J_s-J^d_s) -\frac{\Delta_{\text{Z}}}{2}
+c^2_{\theta_{\alpha}}(s^2_{\theta_{\alpha}}\epsilon_{\text{cap}}
-2 (1+s^2_{\theta_{\alpha}})\epsilon^-_X),\notag\\
\tilde{f}^{\text{c}}&=\frac{\Delta_{\text{Z}}}{2}
+2c^2_{\theta_{\alpha}}\epsilon^-_X  \left(1-\frac{\Delta_{\text{SAS}}^2}{\Delta_{\text{Z}}^2}\right).
\end{align}
From \eqref{limitscantedhamiltonian2-1} we have the dispersion and the coherence length for the mode $\eta^\text{sc}_2$, 
\begin{align}
E^{\eta^\text{sc}_2}&=\frac{4}{\rho_0}J^\alpha_1 \boldsymbol{k}^2+\Delta_{\text{Z}},  \quad
\ \xi^{\eta^\text{sc}_2}=2l_B\sqrt{\frac{\pi J_1^{{\alpha}} }{\Delta_{\text{Z}} }},  
\label{dispersioncoherenceeta2s}
\end{align}
which have exactly the same value as \eqref{dispersioncoherenceeta1s}.

We next analyze ${\mathcal{H}}^{\text{sc}}_{2,2}$ and take $\Delta_{\text{SAS}}=0$ for the sake of the simplicity. 
This Hamiltonian can be diagonalized as
\begin{align}
{\cal{H}}^{\text{sc}}_{2,2}=\left(
\begin{array}{c}
\check{\pi}^{\text{sc}}_{xz,\boldsymbol{k}} \\
\check{\pi}^{\text{sc}}_{yy,\boldsymbol{k}}  \\
\check{\pi}^{\text{sc}}_{yz,\boldsymbol{k}}   \\
\check{\pi}^{\text{sc}}_{xy,\boldsymbol{k}}  \\
\end{array}
\right)^\dagger 
\left(
\begin{array}{cccc}
\lambda^{\text{sc}}_{xz} & 0  & 0 & 0 \\
0 &\lambda^{\text{sc}}_{yy} & 0 & 0 \\
0 & 0 & \lambda^{\text{sc}}_{yz} &  0     \\
0 & 0 & 0 & \lambda^{\text{sc}}_{xy}  \\
\end{array}
\right) 
\left(
\begin{array}{c}
\check{\pi}^{\text{sc}}_{xz,\boldsymbol{k}}  \\
\check{\pi}^{\text{sc}}_{yy,\boldsymbol{k}}  \\
\check{\pi}^{\text{sc}}_{yz,\boldsymbol{k}}   \\
\check{\pi}^{\text{sc}}_{xy,\boldsymbol{k}}  \\
\end{array}
\right), 
\label{canteddiagonalizedhamiltonian} 
\end{align}
where 
\begin{align}
\lambda^{\text{sc}}_{xz}&=\tilde{F}^{\text{c}}+\tilde{f}^{\text{c}}=\frac{2\boldsymbol{k}^2}{\rho_0}J^d_s +\Delta_{\text{Z}}+4c^2_{\theta_{\alpha}}\epsilon^-_X ,
\notag\\
\lambda^{\text{sc}}_{yy}&=\tilde{F}^{\text{c}}-\tilde{f}^{\text{c}}=
\frac{2\boldsymbol{k}^2}{\rho_0}J^d_s, \notag\\
\lambda^{\text{sc}}_{yz}&=\tilde{D}^{\text{c}}+\tilde{d}^{\text{c}}=
\frac{2\boldsymbol{k}^2}{\rho_0}(c^2_{2\theta_\alpha}J^d_s+s^2_{2\theta_\alpha}J_s) +2s^2_{2\theta_\alpha}(\epsilon^-_D-\epsilon^-_X ),  
\notag\\
\lambda^{\text{sc}}_{xy}&=\tilde{D}^{\text{c}}-\tilde{d}^{\text{c}}=
\frac{2\boldsymbol{k}^2}{\rho_0}J^d_s +\Delta_{\text{Z}}+4c^2_{\theta_{\alpha}}\epsilon^-_X ,
\label{eigenvaluesc} 
\end{align}
and 
\begin{align}
\check{\pi}^{\text{sc}}_{xz,\boldsymbol{k}}=
\frac{\tilde{\pi}^{\text{sc}}_{xz,\boldsymbol{k}}+\tilde{\pi}^{\text{sc}}_{yy,\boldsymbol{k}}}{\sqrt{2}}, \quad
\check{\pi}^{\text{sc}}_{yy,\boldsymbol{k}}=
\frac{-\tilde{\pi}^{\text{sc}}_{xz,\boldsymbol{k}}+\tilde{\pi}^{\text{sc}}_{yy,\boldsymbol{k}}}{\sqrt{2}}, \notag\\
\check{\pi}^{\text{sc}}_{yz,\boldsymbol{k}}=
\frac{\tilde{\pi}^{\text{sc}}_{yz,\boldsymbol{k}}+\tilde{\pi}^{\text{sc}}_{xy,\boldsymbol{k}}}{\sqrt{2}}, \quad
\check{\pi}^{\text{sc}}_{xy,\boldsymbol{k}}=
\frac{-\tilde{\pi}^{\text{sc}}_{yz,\boldsymbol{k}}+\tilde{\pi}^{\text{sc}}_{xy,\boldsymbol{k}}}{\sqrt{2}}.
\label{cnewconjugates}
\end{align}
The fields \eqref{cnewconjugates} satisfy the commutation relation
\begin{align}
\left[ \check{\pi}^{\text{sc}}_{xy,\boldsymbol{k}},\check{\pi}^{\text{sc}}_{xz,\boldsymbol{k}^\prime}\right]&=i\delta(\boldsymbol{k}+\boldsymbol{k}^\prime), \
\left[ \check{\pi}^{\text{sc}}_{yz,\boldsymbol{k}},\check{\pi}^{\text{sc}}_{yy,\boldsymbol{k}^\prime}\right]=i\delta(\boldsymbol{k}+\boldsymbol{k}^\prime).
\label{newcantedccr}
\end{align} 
We can rewrite the Hamiltonian \eqref{canteddiagonalizedhamiltonian} as 
\begin{align}
H^\text{sc}_{2,2}&=\int d^2 k\mathcal{H}^\text{sc}_{2,2}
=\int d^2 k \left[
E^{\eta^\text{sc}_{3}}\eta^{\text{sc}\dagger}_{3,\boldsymbol{k}}\eta^\text{sc}_{3,\boldsymbol{k}}
+E^{\eta^\text{sc}_{4}}\eta^{\text{sc}\dagger}_{4,\boldsymbol{k}}\eta^\text{sc}_{4,\boldsymbol{k}}
\right], 
\end{align}
where 
\begin{align}
E^{\eta^\text{sc}_{3}}&=\frac{4\boldsymbol{k}^2}{\rho_0}J^d_s +2\Delta_{\text{Z}}+8c^2_{\theta_{\alpha}}\epsilon^-_X,\notag\\
E^{\eta^\text{sc}_{4}}&=|\boldsymbol{k}|
\sqrt{\frac{8J^d_s}{\rho_0}
\left(
\frac{2\boldsymbol{k}^2}{\rho_0}(c^2_{2\theta_\alpha}J^d_s+s^2_{2\theta_\alpha}J_s) +2s^2_{2\theta_\alpha}(\epsilon^-_D-\epsilon^-_X )
\right)}.
\label{scmixingdispersions}
\end{align}
The annihilation operators $\eta^\text{sc}_{i,\boldsymbol{k}}$ ($i=3,4$) are given by
\begin{align}
\eta^\text{sc}_{3,\boldsymbol{k}}&=\frac{\check{\pi}^{\text{sc}}_{xy,\boldsymbol{k}}+\check{\pi}^{\text{sc}}_{xz,\boldsymbol{k}^\prime}}{\sqrt{2}}, \notag\\
\eta^\text{sc}_{4,\boldsymbol{k}}&=\frac{1}{\sqrt{2}}\left(\left(\frac{\lambda^{\text{sc}}_{yz}}{\lambda^{\text{sc}}_{yy}}\right)^{\frac{1}{4}}\check{\pi}^{\text{sc}}_{yz,\boldsymbol{k}}+i\left(\frac{\lambda^{\text{sc}}_{yy}}{\lambda^{\text{sc}}_{yz}}\right)^{\frac{1}{4}}\check{\pi}^{\text{sc}}_{yy,\boldsymbol{k}^\prime}\right).
\label{scmixingannihilation}
\end{align} 
They satisfy the commutation relation, 
\begin{align}
\left[\eta^\text{sc}_{i,\boldsymbol{k}},\eta^{\text{sc}\dagger}_{j,\boldsymbol{k}^\prime}\right]
=\delta_{ij}\delta(\boldsymbol{k}-\boldsymbol{k}^\prime), 
\end{align}
with $i,j=3,4$.  

We summarize the Goldstone modes in the CAF phase in the limit $\Delta_{\text{SAS}}\rightarrow0$. 
It is to be emphasized that there emerges one gapless mode, $\eta^\text{sc}_{4,\boldsymbol{k}}$,   
reflecting the realization of an exact and its spontaneous breaking of 
a U(1) part of the SU(4) rotational symmetry.  
Furthermore, it has the linear dispersion relation as in (\ref{scmixingdispersions}), 
as leads to a superfluidity associated with this gapless mode.
All other modes have gaps. 

We comment on the existence of the two modes \eqref{cafcoherentmode1} and \eqref{cafcoherentmode2}. 
Their dispersions \eqref{dispersioncoherenceeta1s} and \eqref{dispersioncoherenceeta2s}
are similar to that of the spin wave \eqref{coherencelength1}.
The difference between the dispersion of these two modes and the spin wave is the stiffness dependence. 
\eqref{dispersioncoherenceeta1s} and \eqref{dispersioncoherenceeta2s}
have the stiffness structure of the linear combination of the intralayer stiffness and interlayer stiffness.  
This can be understood because the CAF phase has a layer correlation. 
The other modes are massive due to the Coulomb energy and the Zeeman gap.

\section{Discussion}

\label{sec:IV}

We have presented a systematic method based on the formula \eqref{su4isospin2} 
to investigate the symmetry breaking pattern
and to derive the effective Hamiltonian for the Goldstone modes 
in the $\nu $=2 bilayer QH system.
There are eight Goldstone modes in each phase, 
which are shown to be smoothly transformed one to another across the phase boundary.
In particular, we have analyzed the CAF phase in detail.

The interlayer phase coherence and the Josephson effect are among the most intriguing phenomena in the $\nu=1$ bilayer QH system\cite{Ezawa:2008ae}. 
They are enhanced in the limit $\Delta _{\text{SAS}}\rightarrow 0$.
It is natural to seek for similar phenomena in the $\nu=2$ bilayer QH system.
We may naively expect them to occur in the pseudospin phase.
However, as we have found, almost all electrons are moved to one of the layers in this limit.

This is not the case in the CAF phase, where the electron densities can be controlled arbitrarily in both layers. 
In the CAF phase we have investigated the dispersion relations
and the coherence length in the limit $\Delta _{\text{SAS}}\rightarrow 0$.
Remarkably, we have found one coherent mode whose coherence length diverges. 
Furthermore it has the linear dispersion relation.
It might be responsible to the interlayer phase coherence.

\begin{acknowledgement}      
Y.~Hama  thanks Kazuhiro Watanabe, Tomoki Ozawa, Tetsuo Hatsuda, Kazunori Itakura, Taro Kanao, Takahiro Mikami, 
and Gergely Fejos for useful discussions and comments.
This research was supported in part by JSPS
Research Fellowships for Young Scientists,
and  a Grant-in-Aid for 
Scientific Research  from the Ministry of
Education, Culture, Sports, Science and Technology (MEXT) of Japan
(Nos. 23340067, 24740184, 21540254).  
\end{acknowledgement}   

\appendix 
\renewcommand{\theequation}{A\arabic{equation}}
\setcounter{equation}{0}
\section*{Appendix A}

The special unitary group SU(N) has $(N^2-1)$ generators. According to the standard notation from 
elementary particle physics\cite{Gell-Mann1}, we denote them as $\lambda_A$, $A=1,2,\ldots,N^2-1$,
which are represented by Hermitian, traceless, $N\times N$ matrices,  
and normalize them as
\begin{equation}
\text{Tr}(\lambda_A\lambda_B)=2\delta_{AB}.    
\label{orthogonality1}
\end{equation}
They are characterized by 
\begin{align} 
\left[ \lambda_A,\lambda_B\right]&=2if_{ABC}\lambda_C, \nonumber\\
\{ \lambda_A,\lambda_B\}&=\frac{4}{N}2d_{ABC}\lambda_C,
\end{align}
where $f_{ABC}$ and $d_{ABC}$ are the structure constant of SU(N). 
We have $\lambda_A=\tau_A$ (the Pauli matrix) with $f_{ABC}=\epsilon _{ABC}$ and $d_{ABC}=0$ in the case of SU(2). 
 
This standard representation is not convenient for our purpose because the spin group is 
$\text{SU}(2)\times \text{SU}(2)$ in the bilayer electron system with the four-component electron field as 
$\Psi=(\psi^{\text{f}\uparrow},\psi^{\text{f}\downarrow},\psi^{\text{b}\uparrow},\psi^{\text{b}\downarrow})$. 
Embedding $\text{SU}(2)\times \text{SU}(2)$ into SU(4) we define the spin matrix by 
\begin{equation}
\tau _{a}^{\text{spin}}=\left( 
\begin{array}{cc}
\tau _{a} & 0 \\ 
0 & \tau _{a}
\end{array}
\right) ,  \label{su4base1} 
\end{equation}
where $a=x,y,z$, and the pseudospin matrices by, 
\begin{align}
\tau _{x}^{\text{ppin}}& = & & \left( 
\begin{array}{cc}
0 & \boldsymbol{1}_{2} \\ 
\boldsymbol{1}_{2} & 0
\end{array}
\right) ,\ \ \ \tau _{y}^{\text{ppin}}=\left( 
\begin{array}{cc}
0 & -i\boldsymbol{1}_{2} \\ 
i\boldsymbol{1}_{2} & 0
\end{array}
\right) ,  \notag \\
\tau _{z}^{\text{ppin}}& = & & \left( 
\begin{array}{cc}
\boldsymbol{1}_{2} & 0 \\ 
0 & -\boldsymbol{1}_{2}
\end{array}
\right) ,  \label{su4base2}
\end{align}
where $\boldsymbol{1}_{2}$ is the unit matrix in two dimensions. Nine remaining
matrices are simple products of the spin and pseudospin matrices: 
\begin{align}
\tau _{a}^{\text{spin}}\tau _{x}^{\text{ppin}}& =\left( 
\begin{array}{cc}
0 & \tau _{a} \\ 
\tau _{a} & 0
\end{array}
\right) ,\ \ \ \tau _{a}^{\text{spin}}\tau _{y}^{\text{ppin}}=\left( 
\begin{array}{cc}
0 & -i\tau _{a} \\ 
i\tau _{a} & 0
\end{array}
\right) ,  \notag \\
\tau _{a}^{\text{spin}}\tau _{z}^{\text{ppin}}& =\left( 
\begin{array}{cc}
\tau _{a} & 0 \\ 
0 & -\tau _{a}
\end{array}
\right).  \label{su4base3}
\end{align}
We denote them $T_{a0}\equiv \frac{1}{2}\tau_a^{\text{spin}}$, $T_{0a}\equiv 
\frac{1}{2}\tau_a^{\text{ppin}}$,
$T_{ab}\equiv \frac{1}{2}\tau_a^{\text{spin}}\tau_b^{\text{ppin}}$. They satisfy the normalization condition 
\begin{equation}
\text{Tr}(T_{\mu \nu}T_{\gamma\delta})=\delta_{\mu\gamma}\delta_{\nu\delta},
\label{orthogonality2}
\end{equation}
and the commutation relations  
\begin{equation}
[T_{\mu\nu},T_{\gamma\delta}]=if_{\mu\nu,\gamma\delta,\mu^\prime\nu^\prime}T_{\mu^\prime\nu^\prime},
 \label{su4commutator1}
\end{equation}
where $f_{\mu\nu,\gamma\delta,\mu\prime\nu^\prime}$ is  
the SU(4) structure constants in the basis \eqref{su4base1}-\eqref{su4base3}.    
Greek indices run over $0,x,y,z$. 

\renewcommand{\theequation}{B\arabic{equation}}
\setcounter{equation}{0}
\section*{Appendix B}
We express the rotated isospin fields $\mathcal{I}^{\text{sc}}_{\mu\nu}$ in terms of the eight Goldstone fields $\pi^{\text{sc}}_{\mu\nu}$ up to the second order,
\begin{align}   
\mathcal{I}^{\text{sc}}_{x\mu }&=-\pi^{\text{sc}}_{y\mu}+\mathcal{O}(\pi^2), \ \mathcal{I}^{\text{sc}}_{y\mu }=\pi^{\text{sc}}_{x\mu}+\mathcal{O}(\pi^2),\notag\\ 
\mathcal{I}^{\text{sc}}_{0y}&=\frac{\pi^{\text{sc}}_{xz}\pi^{\text{sc}}_{yx}+\pi^{\text{sc}}_{yx}\pi^{\text{sc}}_{xz}-\pi^{\text{sc}}_{yz}\pi^{\text{sc}}_{xx}-\pi^{\text{sc}}_{xx}\pi^{\text{sc}}_{yz}}{2}+\mathcal{O}(\pi^3),\notag\\
\mathcal{I}^{\text{sc}}_{0z}&=\frac{\pi^{\text{sc}}_{xx}\pi^{\text{sc}}_{yy}+\pi^{\text{sc}}_{yy}\pi^{\text{sc}}_{xx}-\pi^{\text{sc}}_{xy}\pi^{\text{sc}}_{yx}-\pi^{\text{sc}}_{yx}\pi^{\text{sc}}_{xy}}{2}+\mathcal{O}(\pi^3),\notag\\
\mathcal{I}^{\text{sc}}_{zx}&=-\frac{\pi^{\text{sc}}_{yx}\pi^{\text{sc}}_{y0}+\pi^{\text{sc}}_{y0}\pi^{\text{sc}}_{yx}+\pi^{\text{sc}}_{xx}\pi^{\text{sc}}_{x0}+\pi^{\text{sc}}_{x0}\pi^{\text{sc}}_{xx}}{2}+\mathcal{O}(\pi^3),\notag\\
\mathcal{I}^{\text{sc}}_{zy}&=-\frac{\pi^{\text{sc}}_{yy}\pi^{\text{sc}}_{y0}+\pi^{\text{sc}}_{y0}\pi^{\text{sc}}_{yy}+\pi^{\text{sc}}_{xy}\pi^{\text{sc}}_{x0}+\pi^{\text{sc}}_{x0}\pi^{\text{sc}}_{xy}}{2}+\mathcal{O}(\pi^3),\notag\\
\mathcal{I}^{\text{sc}}_{zz}&=-\frac{\pi^{\text{sc}}_{yz}\pi^{\text{sc}}_{y0}+\pi^{\text{sc}}_{y0}\pi^{\text{sc}}_{yz}+\pi^{\text{sc}}_{xz}\pi^{\text{sc}}_{x0}+\pi^{\text{sc}}_{x0}\pi^{\text{sc}}_{xz}}{2}+\mathcal{O}(\pi^3),\notag\\
\mathcal{I}^{\text{sc}}_{0x}&=\frac{\pi^{\text{sc}}_{xy}\pi^{\text{sc}}_{yz}+\pi^{\text{sc}}_{yz}\pi^{\text{sc}}_{xy}-\pi^{\text{sc}}_{yy}\pi^{\text{sc}}_{xz}-\pi^{\text{sc}}_{xz}\pi^{\text{sc}}_{yy}}{2}+\mathcal{O}(\pi^3),\notag\\
\mathcal{I}^{\text{sc}}_{z0}&=1-\sum_{\mu=0,x,y,z}\frac{(\pi^{\text{sc}}_{x\mu })^2+(\pi^{\text{sc}}_{y\mu })^2}{2}+\mathcal{O}(\pi^3). 
\label{cafgoldstoneexpansion}
\end{align}
We next  give the relation between the original isospin field $\mathcal{I}_{\mu\nu}$ and the rotated field $\mathcal{I}^{\text{sc}}_{\mu\nu}$ in the s-coordinate of the CAF phase.
\begin{align}
\mathcal{I}_{0x}&=c_{\theta_\alpha}c_{\theta_\beta}c_{\theta_\delta}\mathcal{I}^{\text{sc}}_{0x}+c_{\theta_\alpha}s_{\theta_\beta}\mathcal{I}^{\text{sc}}_{0z}-s_{\theta_\alpha}s_{\theta_\beta}\mathcal{I}^{\text{sc}}_{xx}
+s_{\theta_\alpha}c_{\theta_\beta}c_{\theta_\delta}\mathcal{I}^{\text{sc}}_{xz}\notag\\
&-c_{\theta_\alpha}c_{\theta_\beta}s_{\theta_\delta}\mathcal{I}^{\text{sc}}_{yy}+s_{\theta_\alpha}c_{\theta_\beta}s_{\theta_\delta}\mathcal{I}^{\text{sc}}_{z0},\notag\\
\mathcal{I}_{0y}&=c_{\theta_\delta}\mathcal{I}^{\text{sc}}_{0y}+s_{\theta_\delta}\mathcal{I}^{\text{sc}}_{yx},\notag\\
\mathcal{I}_{0z}&=-c_{\theta_\alpha}s_{\theta_\beta}c_{\theta_\delta}\mathcal{I}^{\text{sc}}_{0x}+c_{\theta_\alpha}c_{\theta_\beta}\mathcal{I}^{\text{sc}}_{0z}-s_{\theta_\alpha}c_{\theta_\beta}\mathcal{I}^{\text{sc}}_{xx}
-s_{\theta_\alpha}s_{\theta_\beta}c_{\theta_\delta}\mathcal{I}^{\text{sc}}_{xz}\notag\\
&+c_{\theta_\alpha}s_{\theta_\beta}s_{\theta_\delta}\mathcal{I}^{\text{sc}}_{yy}-s_{\theta_\alpha}s_{\theta_\beta}s_{\theta_\delta}\mathcal{I}^{\text{sc}}_{z0},\notag\\
\mathcal{I}_{x0}&=c_{\theta_\delta}\mathcal{I}^{\text{sc}}_{x0}+s_{\theta_\delta}\mathcal{I}^{\text{sc}}_{zz},\notag\\
\mathcal{I}_{xx}&=-s_{\theta_\alpha}s_{\theta_\beta}c_{\theta_\delta}\mathcal{I}^{\text{sc}}_{0x}+s_{\theta_\alpha}c_{\theta_\beta}\mathcal{I}^{\text{sc}}_{0z}
+c_{\theta_\alpha}c_{\theta_\beta}\mathcal{I}^{\text{sc}}_{xx}
+c_{\theta_\alpha}s_{\theta_\beta}c_{\theta_\delta}\mathcal{I}^{\text{sc}}_{xz}\notag\\
&+s_{\theta_\alpha}s_{\theta_\beta}s_{\theta_\delta}\mathcal{I}^{\text{sc}}_{yy}+c_{\theta_\alpha}s_{\theta_\beta}s_{\theta_\delta}\mathcal{I}^{\text{sc}}_{z0},\notag\\
\mathcal{I}_{xy}&=\mathcal{I}^{\text{sc}}_{xy},\notag\\
\mathcal{I}_{xz}&=-s_{\theta_\alpha}c_{\theta_\beta}c_{\theta_\delta}\mathcal{I}^{\text{sc}}_{0x}-s_{\theta_\alpha}s_{\theta_\beta}\mathcal{I}^{\text{sc}}_{0z}
-c_{\theta_\alpha}s_{\theta_\beta}\mathcal{I}^{\text{sc}}_{xx}
+c_{\theta_\alpha}c_{\theta_\beta}c_{\theta_\delta}\mathcal{I}^{\text{sc}}_{xz}
\notag\\
&+s_{\theta_\alpha}c_{\theta_\beta}s_{\theta_\delta}\mathcal{I}^{\text{sc}}_{yy}
+c_{\theta_\alpha}c_{\theta_\beta}s_{\theta_\delta}\mathcal{I}^{\text{sc}}_{z0},\notag\\
\mathcal{I}_{y0}&=c_{\theta_\alpha}\mathcal{I}^{\text{sc}}_{y0}-s_{\theta_\alpha}\mathcal{I}^{\text{sc}}_{zy},\notag\\
\mathcal{I}_{yx}&=-c_{\theta_\beta}s_{\theta_\delta}\mathcal{I}^{\text{sc}}_{0y}
+c_{\theta_\beta}c_{\theta_\delta}\mathcal{I}^{\text{sc}}_{yx}
+s_{\theta_\beta}\mathcal{I}^{\text{sc}}_{yz},\notag\\
\mathcal{I}_{yy}&=c_{\theta_\alpha}s_{\theta_\delta}\mathcal{I}^{\text{sc}}_{0x}+s_{\theta_\alpha}s_{\theta_\delta}\mathcal{I}^{\text{sc}}_{xz}
+c_{\theta_\alpha}c_{\theta_\delta}\mathcal{I}^{\text{sc}}_{yy}
-s_{\theta_\alpha}c_{\theta_\delta}\mathcal{I}^{\text{sc}}_{z0},\notag\\
\mathcal{I}_{yz}&=s_{\theta_\beta}s_{\theta_\delta}\mathcal{I}^{\text{sc}}_{0y}
-s_{\theta_\beta}c_{\theta_\delta}\mathcal{I}^{\text{sc}}_{yx}
+c_{\theta_\beta}\mathcal{I}^{\text{sc}}_{yz},\notag\\
\mathcal{I}_{z0}&=s_{\theta_\alpha}s_{\theta_\delta}\mathcal{I}^{\text{sc}}_{0x}-c_{\theta_\alpha}s_{\theta_\delta}\mathcal{I}^{\text{sc}}_{xz}+s_{\theta_\alpha}c_{\theta_\delta}\mathcal{I}^{\text{sc}}_{yy}+c_{\theta_\alpha}c_{\theta_\delta}\mathcal{I}^{\text{sc}}_{z0},\notag\\
\mathcal{I}_{zx}&=-s_{\theta_\beta}s_{\theta_\delta}\mathcal{I}^{\text{sc}}_{x0}+c_{\theta_\beta}\mathcal{I}^{\text{sc}}_{zx}+s_{\theta_\beta}c_{\theta_\delta}\mathcal{I}^{\text{sc}}_{zz},\notag\\
\mathcal{I}_{zy}&=s_{\theta_\alpha}\mathcal{I}^{\text{sc}}_{y0}+c_{\theta_\alpha}\mathcal{I}^{\text{sc}}_{zy},\notag\\
\mathcal{I}_{zz}&=-c_{\theta_\beta}s_{\theta_\delta}\mathcal{I}^{\text{sc}}_{x0}-s_{\theta_\beta}\mathcal{I}^{\text{sc}}_{zx}+c_{\theta_\beta}c_{\theta_\delta}\mathcal{I}^{\text{sc}}_{zz},
\label{cafisospinrelation}
\end{align}
with \eqref{sincosdefinition}.

\end{document}